\documentclass[twocolumn]{aastex61}

\usepackage[utf8]{inputenc}
\DeclareUnicodeCharacter{2212}{-}
\usepackage{graphicx}
\graphicspath{{./figures/}{./}}
\usepackage{amsmath}
\usepackage{amsfonts}
\usepackage{dcolumn}
\usepackage{multirow}
\usepackage{afterpage}
\usepackage{enumitem}
\setenumerate{itemsep=0mm}
\hypersetup{    
  urlcolor        = {blue},
}

\usepackage{lineno}
\usepackage{soul}

\usepackage{soul} 
\usepackage{amsmath}
\usepackage{amssymb}
\usepackage{xspace}
\usepackage{xifthen}
\usepackage{eso-pic}

\definecolor{forestgreen}{HTML}{228B22}
\definecolor{urlblue}{HTML}{000000}

\newcommand{\reportnum}[2]{
  \AddToShipoutPictureBG*{%
    \AtPageUpperLeft{%
      \hspace{0.75\paperwidth}%
      \raisebox{#1\baselineskip}{%
        \makebox[0pt][l]{\textnormal{#2}}
  }}}%
}

\mathchardef\mhyphen="2D

\newlength{\dhatheight}

\newcommand{\code}[1]{\texttt{#1}\xspace}

\newcommand{\secref}[1]{Section~\ref{sec:#1}}

\newcommand{\tabref}[1]{Table~\ref{tab:#1}}

\newcommand{\figref}[1]{Figure~\ref{fig:#1}}

\newcommand{\eqnref}[1]{Equation~\eqref{eqn:#1}}

\newcommand{\bandvar}[2][]{%
  \ifthenelse{\isempty{#1}}{\var{#2}}{\var{#2\_#1}}%
}

\newcommand{\snana}{\code{SNANA}}
\newcommand{\psnid}{\code{PSNID}}
\newcommand{\var}[1]{\ensuremath{\texttt{\MakeUppercase{#1}}}\xspace}

\providecommand\physrep{\ref@jnl{Phys.~Rep.}}%
\providecommand\apjs{\ref@jnl{ApJS}}%
\providecommand{\jcap}{\ref@jnl{JCAP}}%

\begin{document}

\vspace*{-\headsep}\vspace*{\headheight}
{\footnotesize \hfill FERMILAB-PUB-19-342-AE}\\
\vspace*{-\headsep}\vspace*{\headheight}
{\footnotesize \hfill DES-2018-0408}\\

\title{A DECam search for explosive optical transients associated with IceCube neutrinos}


\correspondingauthor{Robert Morgan and Keith Bechtol}
\email{robert.morgan@wisc.edu}
\email{kbechtol@wisc.edu}

\author[0000-0002-7016-5471]{R.~Morgan}
\affil{Physics Department, University of Wisconsin-Madison, 1150 University Avenue Madison, WI  53706, USA}
\affil{Large Synoptic Survey Telescope Corporation Data Science Fellowship Program}

\author{K.~Bechtol}
\affil{Large Synoptic Survey Telescope, 933 North Cherry Avenue, Tucson, AZ 85721, USA}
\affil{Physics Department, University of Wisconsin-Madison, 1150 University Avenue Madison, WI  53706, USA}

\author[0000-0003-3221-0419]{R.~Kessler}
\affil{Department of Astronomy and Astrophysics, University of Chicago, Chicago, IL 60637, USA}
\affil{Kavli Institute for Cosmological Physics, University of Chicago, Chicago, IL 60637, USA}

\author{M.~Sako}
\affil{Department of Physics and Astronomy, University of Pennsylvania, Philadelphia, PA 19104, USA}

\author{K.~Herner}
\affil{Fermi National Accelerator Laboratory, P. O. Box 500, Batavia, IL 60510, USA}

\author{Z.~Doctor}
\affil{Kavli Institute for Cosmological Physics, University of Chicago, Chicago, IL 60637, USA}

\author{D.~Scolnic}
\affil{Kavli Institute for Cosmological Physics, University of Chicago, Chicago, IL 60637, USA}

\author{I.~Sevilla-Noarbe}
\affil{Centro de Investigaciones Energ\'eticas, Medioambientales y Tecnol\'ogicas (CIEMAT), Madrid, Spain}

\author{A.~Franckowiak}
\affil{Deutsches Elektronen Synchrotron DESY, D-15738 Zeuthen, Germany}

\author{K.~N.~Neilson}
\affil{Department of Physics, Drexel University, 3141 Chestnut St, Philadelphia, PA 19104}

\author{M.~Kowalski}
\affil{Institute of Physics, Humboldt-Universität zu Berlin, Newtonstr. 15, 12489 Berlin, Germany}
\affil{Deutsches Elektronen-Synchrotron, D-15735 Zeuthen, Germany}

\author{A.~Palmese}
\affil{Fermi National Accelerator Laboratory, P. O. Box 500, Batavia, IL 60510, USA}

\author{E.~Swann}
\affil{Institute of Cosmology and Gravitation, University of Portsmouth, Portsmouth, PO1 3FX, UK}

\author{B.~P.~Thomas}
\affil{Institute of Cosmology and Gravitation, University of Portsmouth, Portsmouth, PO1 3FX, UK}

\author[0000-0003-4341-6172]{A.~K.~Vivas}
\affil{Cerro Tololo Inter-American Observatory, National Optical Astronomy Observatory, Casilla 603, La Serena, Chile}

\author[0000-0001-8251-933X]{A.~Drlica-Wagner}
\affil{Fermi National Accelerator Laboratory, P. O. Box 500, Batavia, IL 60510, USA}
\affil{Kavli Institute for Cosmological Physics, University of Chicago, Chicago, IL 60637, USA}

\author{A.~Garcia}
\affil{Brandeis University, Physics Department, 415 South Street, Waltham MA 02453}

\author{D.~Brout}
\affil{Department of Physics and Astronomy, University of Pennsylvania, Philadelphia, PA 19104, USA}

\author{F.~Paz-Chinch\'{o}n}
\affil{National Center for Supercomputing Applications, 1205 West Clark St., Urbana, IL 61801, USA}

\author[0000-0002-7357-0317]{E.~Neilsen}
\affil{Fermi National Accelerator Laboratory, P. O. Box 500, Batavia, IL 60510, USA}

\author[0000-0002-8357-7467]{H.~T.~Diehl}
\affil{Fermi National Accelerator Laboratory, P. O. Box 500, Batavia, IL 60510, USA}

\author[0000-0001-6082-8529]{M.~Soares-Santos}
\affil{Brandeis University, Physics Department, 415 South Street, Waltham MA 02453}


\author{T.~M.~C.~Abbott}
\affil{Cerro Tololo Inter-American Observatory, National Optical Astronomy Observatory, Casilla 603, La Serena, Chile}

\author{S.~Avila}
\affil{Instituto de Fisica Teorica UAM/CSIC, Universidad Autonoma de Madrid, 28049 Madrid, Spain}

\author{E.~Bertin}
\affil{CNRS, UMR 7095, Institut d'Astrophysique de Paris, F-75014, Paris, France}
\affil{Sorbonne Universit\'es, UPMC Univ Paris 06, UMR 7095, Institut d'Astrophysique de Paris, F-75014, Paris, France}

\author{D.~Brooks}
\affil{Department of Physics \& Astronomy, University College London, Gower Street, London, WC1E 6BT, UK}

\author[0000-0002-3304-0733]{E.~Buckley-Geer}
\affil{Fermi National Accelerator Laboratory, P. O. Box 500, Batavia, IL 60510, USA}

\author[0000-0003-3044-5150]{A.~Carnero~Rosell}
\affil{Centro de Investigaciones Energ\'eticas, Medioambientales y Tecnol\'ogicas (CIEMAT), Madrid, Spain}
\affil{Laborat\'orio Interinstitucional de e-Astronomia - LIneA, Rua Gal. Jos\'e Cristino 77, Rio de Janeiro, RJ - 20921-400, Brazil}

\author[0000-0002-4802-3194]{M.~Carrasco~Kind}
\affil{Department of Astronomy, University of Illinois at Urbana-Champaign, 1002 W. Green Street, Urbana, IL 61801, USA}
\affil{National Center for Supercomputing Applications, 1205 West Clark St., Urbana, IL 61801, USA}

\author[0000-0002-3130-0204]{J.~Carretero}
\affil{Institut de F\'{\i}sica d'Altes Energies (IFAE), The Barcelona Institute of Science and Technology, Campus UAB, 08193 Bellaterra (Barcelona) Spain}

\author{R.~Cawthon}
\affil{Physics Department, 2320 Chamberlin Hall, University of Wisconsin-Madison, 1150 University Avenue Madison, WI  53706-1390}

\author{M.~Costanzi}
\affil{Universit\"ats-Sternwarte, Fakult\"at f\"ur Physik, Ludwig-Maximilians Universit\"at M\"unchen, Scheinerstr. 1, 81679 M\"unchen, Germany}

\author[0000-0001-8318-6813]{J.~De~Vicente}
\affil{Centro de Investigaciones Energ\'eticas, Medioambientales y Tecnol\'ogicas (CIEMAT), Madrid, Spain}

\author[0000-0002-0466-3288]{S.~Desai}
\affil{Department of Physics, IIT Hyderabad, Kandi, Telangana 502285, India}

\author{P.~Doel}
\affil{Department of Physics \& Astronomy, University College London, Gower Street, London, WC1E 6BT, UK}

\author{B.~Flaugher}
\affil{Fermi National Accelerator Laboratory, P. O. Box 500, Batavia, IL 60510, USA}

\author{P.~Fosalba}
\affil{Institut d'Estudis Espacials de Catalunya (IEEC), 08034 Barcelona, Spain}
\affil{Institute of Space Sciences (ICE, CSIC),  Campus UAB, Carrer de Can Magrans, s/n,  08193 Barcelona, Spain}

\author[0000-0003-4079-3263]{J.~Frieman}
\affil{Fermi National Accelerator Laboratory, P. O. Box 500, Batavia, IL 60510, USA}
\affil{Kavli Institute for Cosmological Physics, University of Chicago, Chicago, IL 60637, USA}

\author[0000-0002-9370-8360]{J.~Garc\'ia-Bellido}
\affil{Instituto de Fisica Teorica UAM/CSIC, Universidad Autonoma de Madrid, 28049 Madrid, Spain}

\author{E.~Gaztanaga}
\affil{Institut d'Estudis Espacials de Catalunya (IEEC), 08034 Barcelona, Spain}
\affil{Institute of Space Sciences (ICE, CSIC),  Campus UAB, Carrer de Can Magrans, s/n,  08193 Barcelona, Spain}

\author[0000-0001-6942-2736]{D.~W.~Gerdes}
\affil{Department of Astronomy, University of Michigan, Ann Arbor, MI 48109, USA}
\affil{Department of Physics, University of Michigan, Ann Arbor, MI 48109, USA}

\author{D.~Gruen}
\affil{Department of Physics, Stanford University, 382 Via Pueblo Mall, Stanford, CA 94305, USA}
\affil{Kavli Institute for Particle Astrophysics \& Cosmology, P. O. Box 2450, Stanford University, Stanford, CA 94305, USA}
\affil{SLAC National Accelerator Laboratory, Menlo Park, CA 94025, USA}

\author{R.~A.~Gruendl}
\affil{Department of Astronomy, University of Illinois at Urbana-Champaign, 1002 W. Green Street, Urbana, IL 61801, USA}
\affil{National Center for Supercomputing Applications, 1205 West Clark St., Urbana, IL 61801, USA}

\author{J.~Gschwend}
\affil{Laborat\'orio Interinstitucional de e-Astronomia - LIneA, Rua Gal. Jos\'e Cristino 77, Rio de Janeiro, RJ - 20921-400, Brazil}
\affil{Observat\'orio Nacional, Rua Gal. Jos\'e Cristino 77, Rio de Janeiro, RJ - 20921-400, Brazil}

\author[0000-0003-0825-0517]{G.~Gutierrez}
\affil{Fermi National Accelerator Laboratory, P. O. Box 500, Batavia, IL 60510, USA}

\author{D.~L.~Hollowood}
\affil{Santa Cruz Institute for Particle Physics, Santa Cruz, CA 95064, USA}

\author{K.~Honscheid}
\affil{Center for Cosmology and Astro-Particle Physics, The Ohio State University, Columbus, OH 43210, USA}
\affil{Department of Physics, The Ohio State University, Columbus, OH 43210, USA}

\author{D.~J.~James}
\affil{Center for Astrophysics $\vert$ Harvard \& Smithsonian, 60 Garden Street, Cambridge, MA 02138, USA}

\author[0000-0003-2511-0946]{N.~Kuropatkin}
\affil{Fermi National Accelerator Laboratory, P. O. Box 500, Batavia, IL 60510, USA}

\author{M.~Lima}
\affil{Departamento de F\'isica Matem\'atica, Instituto de F\'isica, Universidade de S\~ao Paulo, CP 66318, S\~ao Paulo, SP, 05314-970, Brazil}
\affil{Laborat\'orio Interinstitucional de e-Astronomia - LIneA, Rua Gal. Jos\'e Cristino 77, Rio de Janeiro, RJ - 20921-400, Brazil}

\author{M.~A.~G.~Maia}
\affil{Laborat\'orio Interinstitucional de e-Astronomia - LIneA, Rua Gal. Jos\'e Cristino 77, Rio de Janeiro, RJ - 20921-400, Brazil}
\affil{Observat\'orio Nacional, Rua Gal. Jos\'e Cristino 77, Rio de Janeiro, RJ - 20921-400, Brazil}

\author[0000-0003-0710-9474]{J.~L.~Marshall}
\affil{George P. and Cynthia Woods Mitchell Institute for Fundamental Physics and Astronomy, and Department of Physics and Astronomy, Texas A\&M University, College Station, TX 77843,  USA}

\author[0000-0002-1372-2534]{F.~Menanteau}
\affil{Department of Astronomy, University of Illinois at Urbana-Champaign, 1002 W. Green Street, Urbana, IL 61801, USA}
\affil{National Center for Supercomputing Applications, 1205 West Clark St., Urbana, IL 61801, USA}

\author{C.~J.~Miller}
\affil{Department of Astronomy, University of Michigan, Ann Arbor, MI 48109, USA}
\affil{Department of Physics, University of Michigan, Ann Arbor, MI 48109, USA}

\author[0000-0002-6610-4836]{R.~Miquel}
\affil{Instituci\'o Catalana de Recerca i Estudis Avan\c{c}ats, E-08010 Barcelona, Spain}
\affil{Institut de F\'{\i}sica d'Altes Energies (IFAE), The Barcelona Institute of Science and Technology, Campus UAB, 08193 Bellaterra (Barcelona) Spain}

\author[0000-0002-2598-0514]{A.~A.~Plazas}
\affil{Department of Astrophysical Sciences, Princeton University, Peyton Hall, Princeton, NJ 08544, USA}

\author[0000-0002-9646-8198]{E.~Sanchez}
\affil{Centro de Investigaciones Energ\'eticas, Medioambientales y Tecnol\'ogicas (CIEMAT), Madrid, Spain}

\author{V.~Scarpine}
\affil{Fermi National Accelerator Laboratory, P. O. Box 500, Batavia, IL 60510, USA}

\author{M.~Schubnell}
\affil{Department of Physics, University of Michigan, Ann Arbor, MI 48109, USA}

\author{S.~Serrano}
\affil{Institut d'Estudis Espacials de Catalunya (IEEC), 08034 Barcelona, Spain}
\affil{Institute of Space Sciences (ICE, CSIC),  Campus UAB, Carrer de Can Magrans, s/n,  08193 Barcelona, Spain}

\author[0000-0002-3321-1432]{M.~Smith}
\affil{School of Physics and Astronomy, University of Southampton,  Southampton, SO17 1BJ, UK}

\author[0000-0002-7822-0658]{F.~Sobreira}
\affil{Instituto de F\'isica Gleb Wataghin, Universidade Estadual de Campinas, 13083-859, Campinas, SP, Brazil}
\affil{Laborat\'orio Interinstitucional de e-Astronomia - LIneA, Rua Gal. Jos\'e Cristino 77, Rio de Janeiro, RJ - 20921-400, Brazil}

\author[0000-0002-7047-9358]{E.~Suchyta}
\affil{Computer Science and Mathematics Division, Oak Ridge National Laboratory, Oak Ridge, TN 37831}

\author{M.~E.~C.~Swanson}
\affil{National Center for Supercomputing Applications, 1205 West Clark St., Urbana, IL 61801, USA}

\author[0000-0003-1704-0781]{G.~Tarle}
\affil{Department of Physics, University of Michigan, Ann Arbor, MI 48109, USA}

\author{V.~Vikram}
\affil{Argonne National Laboratory, 9700 South Cass Avenue, Lemont, IL 60439, USA}

\author[0000-0002-7123-8943]{A.~R.~Walker}
\affil{Cerro Tololo Inter-American Observatory, National Optical Astronomy Observatory, Casilla 603, La Serena, Chile}

\author[0000-0002-8282-2010]{J.~Weller}
\affil{Excellence Cluster Origins, Boltzmannstr.\ 2, 85748 Garching, Germany}
\affil{Max Planck Institute for Extraterrestrial Physics, Giessenbachstrasse, 85748 Garching, Germany}
\affil{Universit\"ats-Sternwarte, Fakult\"at f\"ur Physik, Ludwig-Maximilians Universit\"at M\"unchen, Scheinerstr. 1, 81679 M\"unchen, Germany}

\begin{abstract}

In this work, we investigate the likelihood of association between realtime, TeV-PeV energy neutrino alerts from IceCube and optical counterparts in the form of core-collapse supernovae (CC SNe).
The optical follow-up of IceCube alerts requires two main instrumental capabilities: (1) deep imaging, since 73\% of neutrinos would come from CC SNe at redshifts $z > 0.3$, and (2) a large field of view (FoV), since typical IceCube muon neutrino pointing accuracy is on the order of $\sim1$~deg.
With Blanco/DECam ($gri$ to 24th magnitude and $2.2$~deg diameter FoV), we performed a triggered optical follow-up observation of two IceCube alerts, IC170922A and IC171106A on $\sim6$~nights during the $\sim3$~weeks following each alert.
For the IC170922A (IC171106A) follow-up observations, we expect that 12.1\% (9.5\%) of coincident CC SNe at $z \lesssim 0.3$ are detectable, and that on average, 0.23 (0.07) unassociated SNe in the neutrino 90\% containment regions also pass our selection criteria.
We find two candidate CC SNe that are temporally coincident with the neutrino alerts in the FoV, but none in the 90\% containment regions, which is statistically consistent with expected rates of background CC SNe for these observations. 
If CC SNe are the dominant source of TeV-PeV neutrinos, we would expect an excess of coincident CC SNe to be detectable at the $3\sigma$ confidence level using DECam observations similar to those of this work for $\sim60$ ($\sim200$) neutrino alerts with (without) redshift information for all candidates.

\end{abstract}


\section{Introduction}
\label{sec:intro}

The detection of TeV-PeV energy astrophysical neutrinos with the IceCube detector \citep[]{IceCube:2013} has opened a door for multimessenger studies of energetic astrophysical environments \citep[]{Franckowiak:2017}.
Such high-energy neutrinos are generally understood to be produced exclusively by the interaction of hadrons that have been accelerated to high energies \citep[]{PhysRevD.84.082001}.
Neutrinos are largely unaffected by intervening matter and radiation fields and thus can carry information from larger redshifts, and from deeper within opaque sources, than any other particle messenger \citep[]{Chiarusi:2010, Baret:2011}.
These properties make TeV-PeV energy neutrinos informative probes of high-energy environments, with the potential to provide unique insight to explosive events, such as supernovae \citep[]{Gaisser:1987} and active galaxies \citep[]{Silberberg:1979}, across cosmic time. 

In neutrino astronomy, the low neutrino interaction cross section limits the event rate of high-confidence astrophysical neutrinos to a few events per year \citep[]{Aartsen:201730}. 
To enable time-domain searches for counterparts to the detected astrophysical neutrinos, the IceCube Collaboration has implemented a realtime alert system for the highest confidence and best localized neutrino events. 
22 public realtime neutrino alerts have been issued since 2016 \citep[]{IceCube:2018-data}, leading to the identification of the first compelling electromagnetic counterpart of a TeV energy neutrino, the flaring gamma-ray blazar TXS~0506+056 \citep[]{eaat1378, eaat2890}.
We discuss the connection of TXS~0506+056 to this work in Section \ref{sec:data}.
The sources of the other 21 alerts remain unknown.

The observed flux for all astrophysical neutrinos with TeV-PeV energies is nearly isotropic, supporting a primarily extragalactic origin \citep[]{0004-637X-809-1-98, Ahlers:2016}. 
Several prominent non-thermal and extragalactic source classes, including gamma-ray bursts \citep[]{2041-8205-805-1-L5}, gamma-ray blazars \citep[]{0004-637X-835-1-45}, and star-forming galaxies \citep[]{Bechtol:2017}, have been suggested and now have stringent upper limits on their total contributions to the IceCube signal.
Many analyses have proposed that a subset of core-collapse supernovae (CC SNe) have internal jets and/or shocks that produce prompt TeV-PeV neutrino emission \citep[among several others]{Thompson:2003, Razzaque:2004, Ando:2005, Woosley2005, Murase:2013}.

In this work, we investigate the possibility to determine observationally,
whether CC SNe contribute to the TeV-PeV energy neutrino flux via a prompt neutrino emission mechanism.
In order to identify a CC SN associated with a neutrino alert, a follow-up of the event must begin as soon as possible after the neutrino signal has been observed. This should happen ideally within 1-2 days since some explosion models predict a fast-rising electromagnetic flux on the order of days \citep[]{Gonzalez:2015}.
It is also worth noting that the rise-times of SN explosion models are relatively unconstrained, so the precise determination of the explosion time from observations is challenging.
Additionally, the high matter densities in collapsing stars are expected to be opaque to gamma-rays \citep[]{Meszaros:2001, Dermer:2003, Senno:2016, Tamborra:2016}, meaning the neutrino signal would lack an accompanying gamma-ray burst.
SNe, though, are characteristically bright in the optical bands despite the cosmological distances.
Therefore, triggered optical follow-up of the best-localized IceCube neutrino events is an attractive way to search for CC SNe in coincidence with individual neutrino sources \citep[]{Kowalski:2007xb}.
We discuss the physical model on which we base our search and its underlying assumptions in \secref{model}.

For optical follow-up to be feasible, the instrument field of view (FoV) must be matched to the solid angle on the sky containing the neutrino in 90\% of cases (the 90\% confidence-level containment region) such that there is a high probability of capturing the neutrino source without becoming overwhelmed by false positives.
The median angular resolution for neutrinos detected by IceCube with energies above 100 TeV is less than 1 degree \citep[]{0004-637X-835-2-151}.
In addition, the neutrino emission from CC SNe is expected to follow the cosmic star-formation rate, in which case the majority of neutrinos detected at the Earth would be produced by distant sources \citep[]{Strigari:2005}. Therefore, the instrument must also reach an imaging depth sufficient for observing the faint optical signal of the distant CC SNe.
This scenario has motivated several optical follow-up efforts, including programs with the Robotic Optical Transient Search Experiment (ROTSE) and the Palomar Transient Factory (PTF) that reach typical limiting magnitudes of 17 and 21, respectively \citep[]{0004-637X-811-1-52, refId0}. 
The most recent optical follow-up was performed by Pan-STARRS1 \citep[limiting magnitude $\sim22.5$,][]{Kankare:2019} in which none of the transient sources found close in direction to five IceCube alerts had convincing evidence for an association.
One plausible explanation for the lack of neutrino candidates found by these studies is low sensitivity to faint, distant objects.
We elaborate on the need for deep imaging in \secref{discussion}. One of the primary goals of this work is to quantify the sensitivity of optical follow-up campaigns to a TeV-PeV neutrino-emitting CC SNe.
In this work, we present the deepest optical follow-up to date of IceCube alerts and go beyond the scope of previous studies by characterizing the sensitivity of our follow-up campaign via a maximum likelihood analysis of SNe simulations.

On September 22, 2017 and November 6, 2017, we received alerts for individual $\sim200$~TeV neutrinos detected by IceCube with good localization and high probability to be of astrophysical origin (IC170922A and IC171106A, respectively).
We used the Dark Energy Camera (DECam) mounted on the Blanco $4 \textrm{-m}$ telescope in Chile to perform triggered optical follow-up of the two IceCube neutrino alerts. 
Blanco/DECam’s $2.2 \textrm{ deg}$ diameter field of view was able to cover the entire 90\% confidence level containment regions in a single pointing for each alert and reached a limiting $r$-band magnitude of 23.6 mag for a $5\sigma$ detection, allowing for a higher sensitivity to CC SNe out to larger redshifts compared to previous efforts.
The details of our follow-up observations are presented in \secref{data}.

As a result of the wide field of view and imaging depth, DECam is capable of finding several hundred transient objects per follow-up.
Therefore, to expedite the screening procedure and standardize selection methodology, in \secref{sensitivity} we present an automated candidate selection pipeline for CC SNe exploding at the time of the neutrino alert. 
We apply the pipeline to our observations and present the results in \secref{data_results}.
In \secref{discussion} we supplement our pipeline with calculations of the likelihood of association between a neutrino alert and likely CC SNe selected by the pipeline. We also describe the requirements of an optical follow-up campaign capable of determining whether CC SNe contribute to the TeV-PeV IceCube neutrino flux at a high confidence level.
We conclude in \secref{conclusion}.

\section{Physical Model}
\label{sec:model}
In this analysis, we evaluate the likelihood that TeV-PeV energy neutrinos detected by IceCube are created by a prompt emission mechanism during the collapse. 
We take this mechanism to be a relativistic jet within the collapsing massive star \citep[]{refId0} such that hadrons within the star are boosted to TeV-PeV energies. 
The boost could be caused by several factors, such as shock-breakout or the espresso mechanism \citep[]{2041-8205-811-2-L38}.
Within the collapsing star, boosted protons could interact with photons, leading to the production of charged pions which would decay to produce energetic neutrinos \citep[]{Gaisser:1987}. The high-energy neutrinos then 
escape the collapsing star due to their low interaction cross section \citep[]{1009-9271-6-S1-23}. 
We make the assumption that the direction of the internal relativistic jet has no influence on our ability to detect the optical signal of the explosion.

With the above model for the explosion, the multimessenger signal detectable on Earth would consist of the TeV-PeV energy neutrino, and the electromagnetic signature of a CC SNe.
The neutrino signal could be an individual neutrino or several neutrinos arriving close in time, a so-called neutrino multiplet.
A multiplet indicates a closer source, since only in that case is the mean number of expected neutrino events detected by IceCube expected to be larger than one. 
The majority of singular neutrino events comes from more distant sources with mean number of expected neutrino events much smaller than one \citep[]{multiplet, Strotjohann:2018ufz}.
In this analysis, the alerts we followed-up are single-neutrino alerts.
The neutrino signal is expected to be detected on Earth a few hours to $\sim1$ day before the electromagnetic signal becomes bright enough to be detectable.
This time delay is a direct result of our assumed explosion mechanism: the neutrinos are emitted at the beginning of the explosion and nearly immediately escape the star, while photons are emitted throughout the typically week-scale process and have a much shorter mean-free-path between interactions inside the star.
Furthermore, gamma rays created by the prompt neutrino mechanism have a high probability of being absorbed by the dense stellar material, so the dominant electromagnetic signal would be seen in the optical wavelength range with a delay of a few hours to $\sim1$ day.

To test our hypothesis that CC SNe contribute to the TeV-PeV energy neutrino flux detected by IceCube, we perform a maximum likelihood analysis of the objects found by Blanco/DECam during our triggered follow-up observations. 
For this analysis, we derive the expected redshift distributions of background SNe and associated CC SNe in \secref{sensitivity}. 
Our derivation of the signal sample redshift distribution is based on three assumptions: the IceCube Collaboration accurately reports the probability that an observed neutrino is astrophysical (i.e. not created by cosmic rays interacting in the atmosphere), CC SNe are the sole component of the TeV-PeV energy neutrino flux, and CC SN redshifts are distributed according to the cosmic massive star formation rate \citep[]{Madau:2014}. 
We elaborate on our first assumption in \secref{icecube}. 
The second assumption is relaxed in \secref{discussion} when we assess the necessities of a sustained optical follow-up campaign for different fractions of CC SNe contribution to the IceCube flux. 
Lastly, the third assumption is physically motivated since stars with mass greater than $\sim8 M_\odot$ are expected to end in CC SNe \citep[]{1995ApJ...450..830B}.

Based on the final assumption, we obtain the expected redshift  distribution of associated CC SNe that we detect in our follow-up observations using the following equation for the cumulative neutrino intensity as a function of redshift:

\begin{multline} \label{eqn:redshift}
\frac{dN_\nu(E_\nu)}{dE_\nu d\Omega dt_{obs}dA} = \\ 
\int^{z_\textrm{max}}_0 \frac{dV}{dzd\Omega}\frac{1}{4\pi D^2_L}\frac{dN_\nu (E_\nu (1+z))}{dE_\nu}\frac{1}{1+z}\mathcal{R}(z)\varepsilon(z)dz.
\end{multline}

In Equation \ref{eqn:redshift}, $dV / (dzd\Omega) $ is the co-moving volume element, $1 / (4\pi D^2_L)$ is a division by the surface area of a sphere with radius equal to the luminosity distance, $dN_\nu (E_\nu (1+z)) / dE_\nu$ is the predicted number of observed neutrinos with energy $E_\nu$ from a CC SN at a given redshift, $1 / (1+z)$ is the general relativistic time-dilation factor, $\mathcal{R}(z)$ is the formation rate for stars that will become CC SNe at a given redshift, and $\varepsilon(z)$ is the optical detection efficiency for associated CC SNe.
Integrating from $z=0$ to $z=z_\textrm{max}$ gives the cumulative neutrino intensity from CC SNe which explode up to redshift $z_\textrm{max}$.
For this calculation and for the simulations used in \secref{sensitivity}, we adopt a flat $\Lambda$CDM cosmology with Hubble constant $H_0 = 67.77$ km s$^{-1}$ Mpc$^{-1}$ following the type Ia SNe measurement made by the Dark Energy Survey \citep[]{DESSN}. 
We also take fractions of the universe's energy density made up by matter and dark energy to be $\Omega_m = 0.298$ and $\Omega_\Lambda = 0.702$, respectively, following the combined probe measurements of \citep[]{PhysRevD.98.043526}.

\section{Observations and Data}
\label{sec:data}

In this section, we provide background on the  instruments used in this analysis and detail the relevant observations. We summarize this section in \tabref{Alerts}.

\begin{table}
\centering
\begin{tabular}{|l|c|c|}
\hline\hline
\multicolumn{3}{c}{\textit{IceCube Neutrino Alerts}} \\
\hline\hline
Name & IC170922A & IC171106A \\
\hline
Date & 22 Sep 2017 & 6 Nov 2017 \\
\hline
Time (UTC) & 20:54:30.43 & 18:39:39.21 \\
\hline
Neutrino Energy (TeV) & 120.0 & 230.0 \\
\hline
RA (deg) & $77.43^{+0.95}_{-0.65}$ & $340.25^{+0.70}_{-0.25}$ \\
\hline
Dec (deg) & $+5.72^{+0.70}_{-0.50}$ & $+7.31^{+0.35}_{-0.25}$ \\
\hline
Containment Region & \multirow{2}{*}{0.97} &  \multirow{2}{*}{0.57} \\[-0.15cm]
Area (90\% C.L.) (deg$^2$) & &\\
\hline
Signalness & 0.56507 & 0.74593 \\
\hline \hline
\multicolumn{3}{c}{\textit{Blanco/DECam Follow-Ups}} \\
\hline \hline
Observing Nights & \multirow{2}{*}{1, 3, 10, 20, 21} & \multirow{2}{*}{1, 2, 4, 10, 16} \\[-0.15cm]
after Alert & & \\
\hline
First Night - Alert & \multirow{2}{*}{10.72} & \multirow{2}{*}{7.22} \\[-0.15cm]
Time Difference (hours) & & \\
\hline
Filters & \textit{g r i}$^\textrm{a}$ & \textit{g r i z}$^\textrm{b}$ \\
\hline
Exposure Time (s) & \multirow{2}{*}{2 $\times$ 150} & \multirow{2}{*}{2 $\times$ 150} \\[-0.15cm]
per Band per Epoch & & \\
\hline
Effective DECam  & \multirow{2}{*}{3.12} & \multirow{2}{*}{2.71} \\[-0.15cm]
FoV Area (deg$^2$) & & \\
\hline
$5\sigma$ Mag Limit ($g,r,i$)  & 23.6, 23.7, 23.3 & 23.4, 23.5, 23.1 \\
\hline 
Average Airmass & 1.24 & 1.26 \\
\hline \hline 
\end{tabular}
\caption{\textit{Top}: Properties of the two southern-sky, high-energy IceCube alerts observable from CTIO from Aug 2017 through Jan 2018. ``Signalness'' is described in Section 2.1. \textit{Bottom}: DECam observing cadences and conditions. \\
$^\textrm{a}$ Only $gr$ were used on the final observing night. \\ $^\textrm{b}$ $gri$ for all nights except the final night, which was $riz$. 
}
\label{tab:Alerts}
\end{table}

\subsection{IceCube Real-time Neutrino Alerts}
\label{sec:icecube}


IceCube is a cubic-kilometer scale neutrino detector located at the geographic South Pole and built into the Antarctic ice at depths ranging from $1450 \textrm{ m}$ to $2450 \textrm{ m}$ \citep[]{icecube_detector}. 
Incident neutrinos are detected indirectly through Cherenkov radiation emitted by secondary
charged particles produced in the interaction of a neutrino with a nucleus of the ice.
Photomultipliers in 5160 optical modules sense the Cherenkov light.
Events detected by IceCube are classified as ``shower'' events or ``track'' events. 
Showers are produced in charged-current interactions of electron and tau neutrinos and in neutral-current interactions of all neutrino flavors. 
A shower of secondary particles with a few meters extension are produced, which is small compared to the distance between the optical modules. 
Therefore, the Cherenkov light produced by the shower particles appears almost spherical.
Such events can be reconstructed with an angular resolution of the incoming neutrino on the order of $\sim15$~deg. 
Tracks are produced by charged-current interactions of muon neutrinos in or close to the detector, which produce a muon that travels along a long linear path through the ice, providing a good lever arm for the angular reconstruction. 
Specifically, for track neutrino events with energies above 100~TeV, IceCube achieves median angular resolutions for high-energy starting events (HESE) and extremely high energy events (EHE) of 0.4-1.6~deg and $<$ 1.0~deg, respectively
\citep[]{1748-0221-9-03-P03009, Aartsen:201730}. 

In an effort to locate the sources of these neutrinos and to facilitate follow-up studies, the IceCube Collaboration has created a real-time alert system for events of likely astrophysical origin \citep[]{Aartsen:201730}. 
The system has a median latency of 33 seconds in which neutrino events are processed and 
reconstructed.
Within three minutes, alerts are published to streams such as the Astrophysical Multimessenger Observatory Network \citep[AMON,][]{SMITH201356, 2017arXiv170804724K} and the Gamma-ray Coordination Network \citep[GCN]{GCN:1998}.
Alerts are accompanied by an estimate of the probability that the neutrino is of astrophysical origin, referred to as the ``signalness''.
The signalness describes how likely the event is of astrophysical origin relative to the total atmospheric background rate and is a function of the neutrino energy and the zenith angle.
For more information on the calculation of the signalness, refer to the discussion in \cite{Aartsen:201730}.

\begin{figure}
\includegraphics[width=\columnwidth]{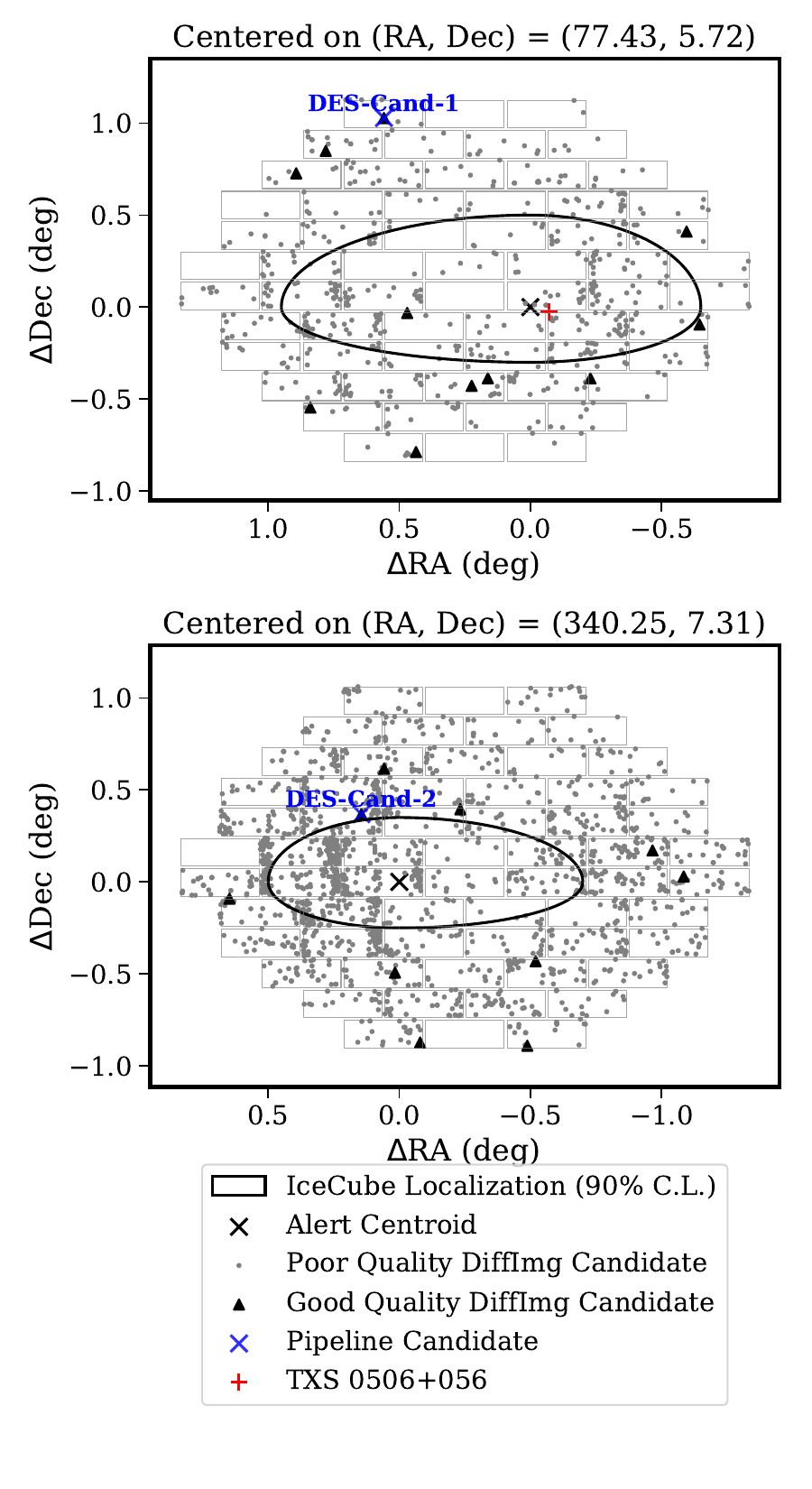}
\caption{The DECam fields of view for the IC170922A follow-up (\textit{top}) and for the IC171106A follow-up (\textit{bottom}) showing difference imaging products, the containment region for the neutrino direction at the 90\% confidence level, and the candidates selected by our Neutrino Candidate Identification Pipeline (\texttt{NCIP}). The location of TXS 0506+056 is also shown for reference. 
}
\label{fig:Localization}
\end{figure}

\begin{figure}
\includegraphics[width=\columnwidth]{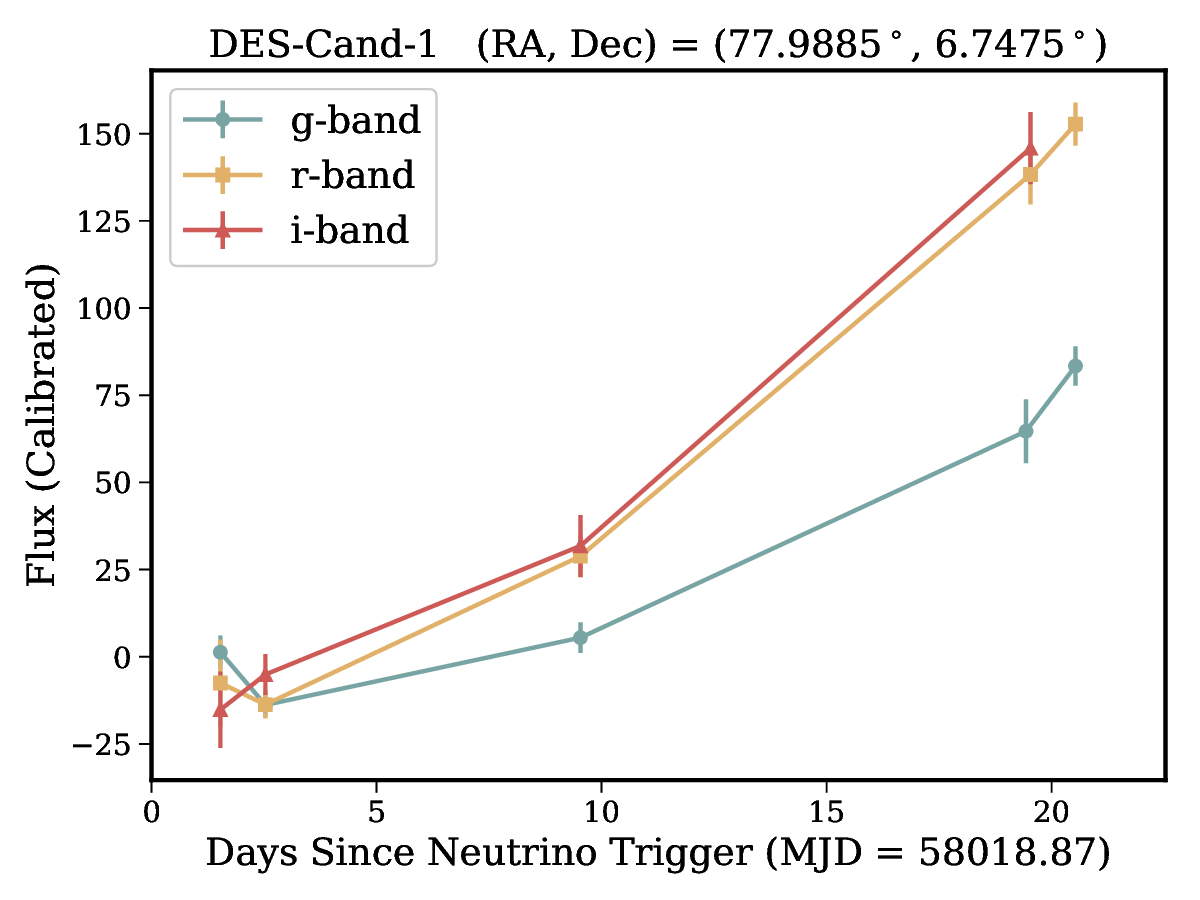}
\includegraphics[width=\columnwidth]{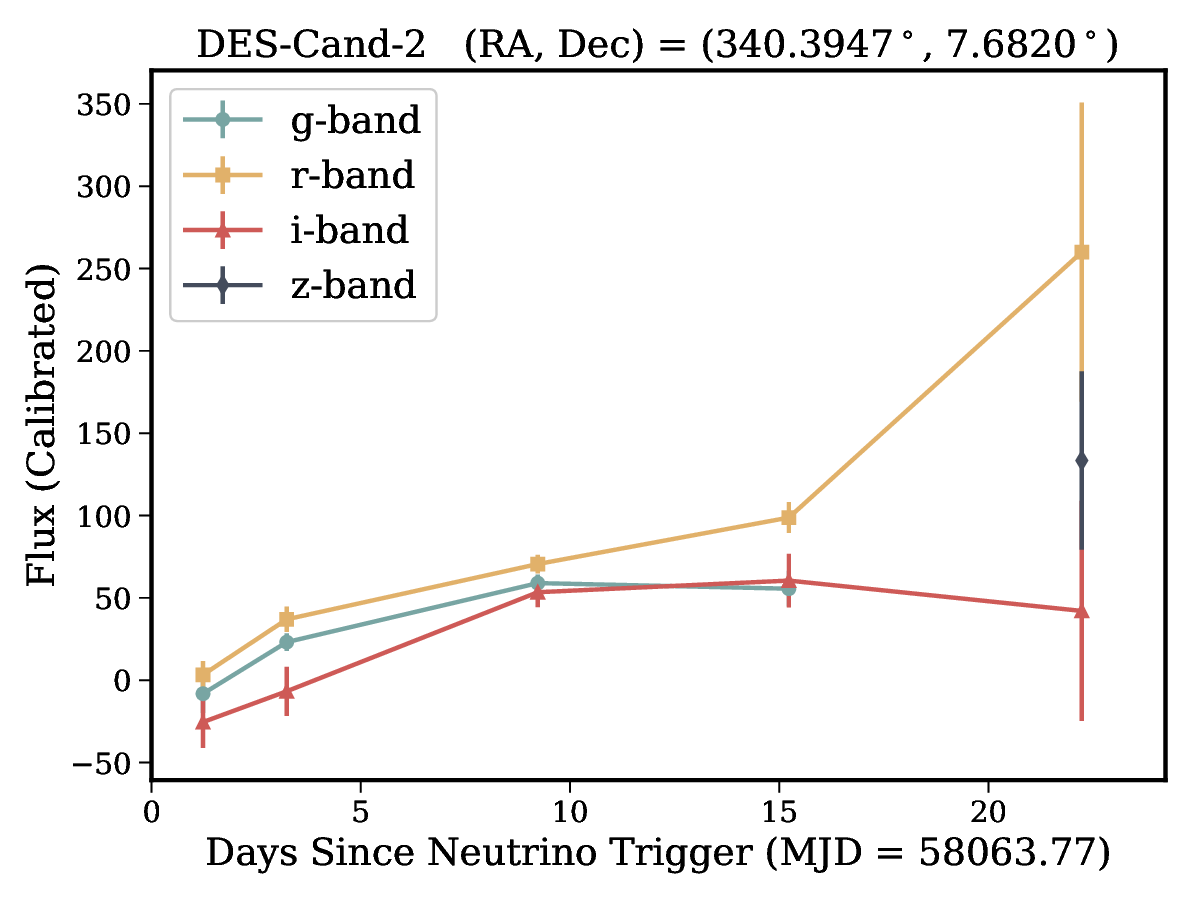}
\caption{\textit{Top}: Light curves in $gri$ for DES-Cand-1, one of the difference imaging products selected by \texttt{NCIP} in the IC170922A follow-up. \textit{Bottom}: Light curves in $griz$ for DES-Cand-2, the candidate from the IC171106A follow-up.}
\label{fig:LightCurve}
\end{figure}

On September 22, 2017 IceCube detected a singular high-energy neutrino. 
Another singular high-energy neutrino was detected on November 6, 2017. 
Both events had good angular resolution and are likely to be of astrophysical origin.
The details of the AMON alerts are presented in the top portion of \tabref{Alerts}. 
Both alerts were track-type singular neutrinos with energies on the 100-200 TeV scale. 
These alerts featured signalness scores of $\sim0.57$ and $\sim0.75$ and  90\% confidence level containment region areas smaller than the field of view of DECam.
The high energies and signalness scores suggest an astrophysical origin of the events and motivate the search for an optical counterpart in form of CC SNe.

Of the 31 alerts issued thus far by the IceCube Collaboration, 21 have occurred after the start of this analysis. 
Of the 21 possible alerts, 16 were either deemed unobservable based on atmospheric, moon, or sun conditions, or were retracted by the IceCube Collaboration. 
The two alerts presented here are two of only five alerts that have been observable from the location of the Blanco Telescope in Chile. 
The additional alerts (IC181023A, IC190331A, and IC190503A) were observed and will be discussed in a future work.

\subsection{DECam Imaging Follow-up}
The Dark Energy Camera \citep[]{Flaugher:2015} is a 570-Megapixel optical imager mounted on the 4-m Blanco Telescope at Cerro Tololo Inter-American Observatory (CTIO) in Chile.
Blanco/DECam's location, field of view ($\sim3 \textrm{ deg}^2$), and deep imaging capabilities 
make it the only southern hemisphere imager with a wide field of view matched to the IceCube angular resolution and a large enough aperture to efficiently detect explosive optical transients at the redshifts relevant for proposed neutrino sources.

The Dark Energy Survey (DES) is a wide-field optical survey (expected to reach a $10\sigma$ depth for point sources of $grizY$ = 25.2, 24.8, 24.0, 23.4, 21.7 mag over 5,000 deg$^2$) with 577 full DECam observing nights \citep[]{des_cal}.
Operating within a large survey program is ideal for our follow-up study because there is no need to interrupt community observers, and the interruptions are short (~20 min).
Specifically, we use triggered target-of-opportunity (ToO) observations to promptly respond to IceCube alerts.
Our observing strategy was to dedicate $\sim1$~month to characterize the rise and peak of potentially-associated CC SNe for each alert.
During the $\sim1$~month observing period, we took observations roughly every 5-6 nights depending on atmospheric and moon conditions.
As well, we added an additional observing night to the start of the follow-up to look for rapidly-evolving optical transients.
Observations consisted of two $150$~sec exposures  in the each of the $gri$ optical bands, and exposures within the same band were stacked to reduce noise and increase imaging depth.
For the IC170922A follow-up, we also adjusted the pointing of the telescope by 0.01 deg in both RA and DEC between exposures of the same band to cover areas of the sky that mapped to gaps between adjacent CCDs, leading to a slightly larger effective DECam FoV area in Table \ref{tab:Alerts}.


The DECam observations were processed by the DES Difference Imaging Pipeline \citep[\code{DiffImg}]{1538-3881-150-6-172}, which subtracts template images from the observations and produces catalog-level photometric fluxes in each band for all detected transients in the field of view.
For this analysis, since our observations did not overlap with the DES footprint where template images would be readily available, we use the first night images as templates.
Observations are also run through \texttt{autoscan} \citep[]{autoscan}, which identifies potential image artifacts and poor image subtractions.
This process injects fake artifacts into the images to train the algorithm to remove artifacts that may be present in the images. 
The probability that a \code{DiffImg} candidate is a real astrophysical object as opposed to an artifact is characterized by the \texttt{autoscan} Machine-Learning (ML) score.
This score is utilized in \secref{sensitivity}.
From the \code{DiffImg} outputs, objects that are likely CC SNe and temporally coincident with the neutrino alert based on light curve properties are selected by an automated neutrino source candidate identification pipeline described in \secref{sensitivity}.
The \code{DiffImg} products and pipeline candidates for both IC170922A and IC171106A are shown in \figref{Localization}. The light curves of the pipeline candidates are shown in \figref{LightCurve}.







\subsubsection{DECam Observations of TXS 0506+056}

One of the two neutrino alerts considered in this work, IC170922A, is likely to be associated with the flaring gamma-ray blazar TXS 0506+056.
The link is based on an analysis of nearly a decade of \textit{Fermi}-LAT observations of gamma-ray blazars, as well as triggered observations of TXS 0506+056 during its flaring state covering the full electromagnetic spectrum \citep[]{eaat1378}.
In addition, a search in archival IceCube data at the position of TXS 0506+056 revealed an excess of lower energy neutrinos during a 158-day period in 2014 and 2015 with a significance of 3.5 $\sigma$ \citep[]{eaat2890}.

We mark the location of TXS 0506+056 within the DECam field of view in Figure \ref{fig:Localization}.
TXS 0506+056 was bright in optical bands around the time of the neutrino alert, reaching $V \sim 14 \textrm{ mag}$ \citep[]{eaat1378}. 
Our choice of $150$~sec exposure times was optimized to search for faint optical transients at $\sim23$~mag, and as a result, TXS 0506+056 is saturated in the DECam images due to the finite dynamic range of the camera.
Optical imaging, spectroscopy, and polarimetry for the event were obtained from combination of ASAS-SN, Kanata/HONIR, Kiso/KWFC,  Liverpool Telescope, and Subaru/FOCAS \citep[]{eaat1378}.

In the discussion that follows, we quantify the sensitivity of DECam to explosive optical transients independent of the probable association between IC170922A and TXS 0506+056.

\section{Sensitivity Analysis Using Simulations}
\label{sec:sensitivity}

\begin{table*}[t]
  \centering
  \begin{tabular}{|c|cc|cc| D{,}{\pm}{-1} D{,}{ \pm }{-1}|c|c|} \hline  
    \multicolumn{9}{|c|}{IC170922A Follow-up}  \\
    \hline
    \multirow{2}{*}{Sample} & \multicolumn{2}{c|}{Signal Efficiency } & \multicolumn{2}{c|}{Signal Efficiency} & \multicolumn{2}{c|}{\multirow{2}{*}{Background Events$^\textrm{a}$ in IC90}} & \multicolumn{1}{c|}{Data$^\textrm{b}$} & \multicolumn{1}{c|}{Data$^\textrm{b}$} \\[-0.15cm]
     & \multicolumn{2}{c|}{All SNe$^\textrm{a}$} & \multicolumn{2}{c|}{Optically Detectable SNe$^\textrm{a}$} & & & in FoV & in IC90 \\
    \hline
    Cut & $z<0.3$ & $z>0.3$ & $z<0.3$ & $z>0.3$ & \multicolumn{2}{l|}{\hspace{0.74cm} $z<0.3$ \hspace{1.17cm} $z>0.3$} & -- & --\\
    \hline 
    Total & $1.0000_{-0.0001}$ & $1.00000_{-0.00001}$ & --  & -- & 9.26\hspace{0.1cm},\hspace{0.1cm}0.14 & 414.2\hspace{0.1cm},\hspace{0.1cm} 0.9 & 617 & 240\\      
    Quality & $0.199^{+0.004}_{-0.004}$ & $2.47^{+0.16}_{-0.15} \times 10^{-3}$ & $1.0000_{-0.0006}$  & $1.000_{-0.005}$ & 1.48\hspace{0.1cm},\hspace{0.1cm}0.05 & 3.96\hspace{0.1cm},\hspace{0.1cm}0.09 & 11 & 2 \\      
    Rising & $0.198^{+0.004}_{-0.004}$ &  $2.47^{+0.15}_{-0.15} \times 10^{-3}$ & $0.9984^{+0.0007}_{-0.0011}$  & $1.000_{-0.005}$ & 1.00\hspace{0.1cm},\hspace{0.1cm}0.04 & 3.07\hspace{0.1cm},\hspace{0.1cm}0.08 & 10 & 2\\      
    Phase & $0.179^{+0.004}_{-0.004}$ & $2.47^{+0.15}_{-0.15} \times 10^{-3}$ & $0.9964^{+0.0012}_{-0.0015}$  & $1.000_{-0.005}$ & 0.51\hspace{0.1cm},\hspace{0.1cm}0.03 & 1.25\hspace{0.1cm},\hspace{0.1cm}0.05 & 2 & 0\\      
    Classify & $0.121^{+0.003}_{-0.003}$ & $1.26^{+0.11}_{-0.11}\times 10^{-3}$ & $0.608^{+0.011}_{-0.011}$  & $0.51^{+0.03}_{-0.03}$ & 0.116\hspace{0.1cm},\hspace{0.1cm}0.015 & 0.113\hspace{0.1cm},\hspace{0.1cm}0.015 & 1 & 0\\      
    \hline \hline 
     \multicolumn{9}{|c|}{IC171106A Follow-up} \\
    \hline
     \multirow{2}{*}{Sample} & \multicolumn{2}{c|}{Signal Efficiency } & \multicolumn{2}{c|}{Signal Efficiency} & \multicolumn{2}{c|}{\multirow{2}{*}{Background Events$^\textrm{a}$ in IC90}} & \multicolumn{1}{c|}{Data$^\textrm{b}$} & \multicolumn{1}{c|}{Data$^\textrm{b}$} \\[-0.15cm]
     & \multicolumn{2}{c|}{All SNe$^\textrm{a}$} & \multicolumn{2}{c|}{Optically Detectable SNe$^\textrm{a}$} & & & in Fov & in IC90 \\
    \hline
    Cut & $z<0.3$ & $z>0.3$ & $z<0.3$ & $z>0.3$ & \multicolumn{2}{l|}{\hspace{0.74cm} $z<0.3$ \hspace{1.17cm} $z>0.3$} & -- & -- \\
    \hline 
    Total  & $1.0000_{-0.0001}$ & $1.00000_{-0.00001}$ & --  & -- & 5.13\hspace{0.1cm},\hspace{0.1cm}0.10 & 236.4\hspace{0.1cm},\hspace{0.1cm}0.7 & 1868 & 646 \\      
    Quality & $0.134^{+0.003}_{-0.003}$ & $7.1^{+0.9}_{-0.8}\times 10^{-4}$ & $1.0000_{-0.0008}$ & $1.000_{-0.016}$ & 0.65\hspace{0.1cm},\hspace{0.1cm}0.04 & 1.46\hspace{0.1cm},\hspace{0.1cm}0.05 & 10 & 0 \\      
    Rising  & $0.134^{+0.003}_{-0.003}$ & $7.1^{+0.9}_{-0.8}\times 10^{-4}$ & $0.9978^{+0.0010}_{-0.0016}$ & $1.000_{-0.016}$ & 0.43\hspace{0.1cm},\hspace{0.1cm}0.03  & 1.11\hspace{0.1cm},\hspace{0.1cm}0.05 & 8 & 0 \\      
    Phase  & $0.130^{+0.003}_{-0.003}$ & $7.1^{+0.9}_{-0.8}\times 10^{-4}$ & $0.970^{+0.004}_{-0.005}$ & $0.986^{+0.010}_{-0.02}$ & 0.187\hspace{0.1cm},\hspace{0.1cm}0.019 & 0.39\hspace{0.1cm},\hspace{0.1cm}0.03 & 2 & 0 \\      
    Classify  & $0.095^{+0.003}_{-0.003}$ & $3.3^{+0.4}_{-0.4}\times 10^{-4}$ & $0.713^{+0.012}_{-0.012}$ & $0.46^{+0.06}_{-0.06}$ & 0.047\hspace{0.1cm},\hspace{0.1cm}0.010 & 0.020\hspace{0.1cm},\hspace{0.1cm}0.006 & 1 & 0 \\      
    \hline
  \end{tabular}
  \caption{Application of the candidate selection pipeline to simulations and data for IC170922A and IC171106A. The term ``optically detectable'' is equivalent to the light curve passing the quality cut. Background event values have a statistical uncertainty of $\pm \sqrt[]{N/500}$ on the mean number of events generated, corresponding to the number of DECam FoVs simulated. The uncertainties reported for the signal efficiencies give a 68\% confidence interval. The numbers of background events reported are scaled to reflect just the events within the IceCube 90\% CL containment region. The data in the total row is not expected to match the simulations because it can contain non-SNe transients or bad image subtractions that are removed by the quality cut.\\$^\textrm{a}$ Results presented are based on \snana simulations.  \\$^\textrm{b}$ Results presented are the number of remaining candidates from DECam observations. 
}
  \label{tab:CutEfficiencies}
\end{table*}

As evidenced by \figref{Localization}, \texttt{DiffImg} recovers many moving objects, variable objects, and transients within a typical neutrino localization region including asteroids, active galactic nuclei (AGN), and background SNe. 
The background SNe population is composed of type Ia SNe, CC SNe exploding before the neutrino alert but persisting into the observing window, CC SNe exploding near the time of the neutrino alert, and CC SNe exploding after the alert but becoming bright enough to be optically detectable before the observing window has concluded. 
The removal of unassociated transients from our observations is critical for isolating potential neutrino source candidates. 
We therefore develop selection criteria to remove CC SNe exploding before or much later than the neutrino trigger and type-Ia SNe altogether.
The selection criteria applied to the \texttt{DiffImg} output are collectively named the Neutrino Candidate Identification Pipeline (\texttt{NCIP}).
The following analysis demonstrates the extent to which these unassociated transients can be removed from DECam observations while maintaining a high detection efficiency for the potentially associated CC SN.

\subsection{Simulations and Candidate Selection}

\begin{figure*}
\includegraphics[width=\textwidth]{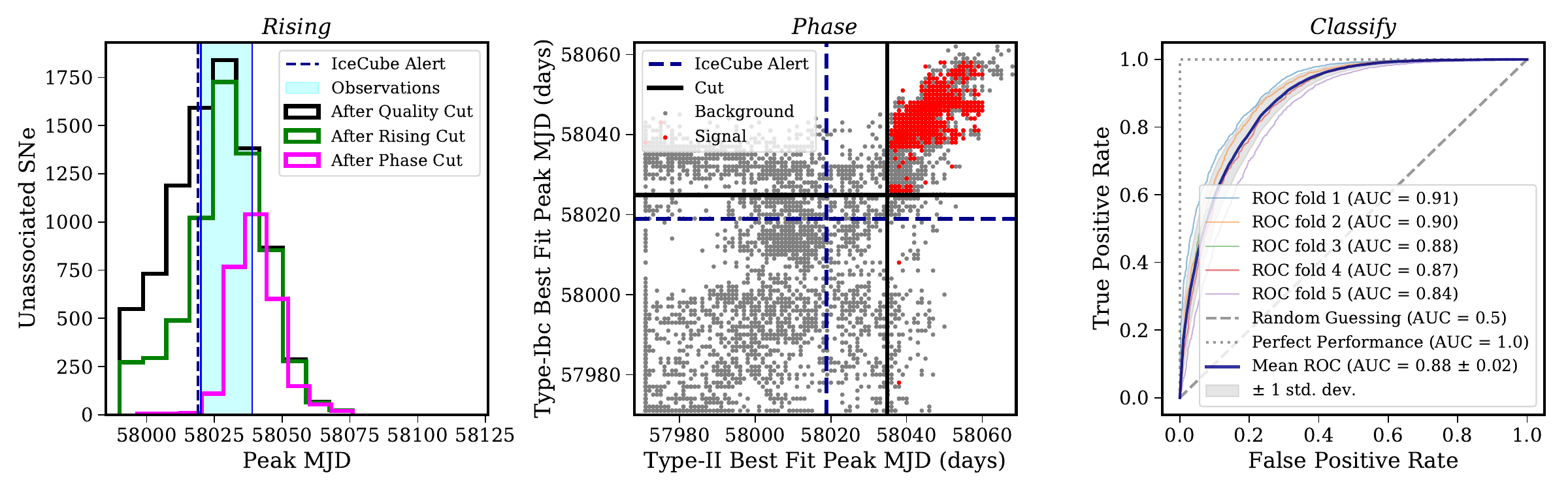}
\caption{Selection criteria motivation and effectiveness based on simulations of the IC170922A follow-up. \textit{Left}: Distribution of peak MJD values from the background sample after successive cuts. \textit{Center}: Best fit peak MJD values from type II and type Ibc template fits with cuts displayed. \textit{Right}: Receiver Operating Characteristic (ROC) curves of our Random Forest Classifier implemented on each fold of the training sample during a 5-fold cross validation. Each curve has an associated Area Under Curve (AUC) to characterize the performance of the classifier.}
\label{fig:selection_criteria}
\end{figure*}

Selecting CC SNe associated with IceCube neutrino alerts requires extrapolating the explosion time of the SN from its light curve.
To model the rise times of SNe, we employ the SuperNova ANAlysis software suite \citep[\snana,][]{Kessler:2009}, which enables us to simulate SN light curves based on our observing cadence. 
Based on \snana simulated SN light curves for our observations, we develop selection criteria to filter out unassociated SNe.
Then by applying the selection criteria to the same simulations, we determine the detection efficiency for an associated CC SN and the expected number of unassociated SNe found by \texttt{NCIP} for each follow-up.




We simulated SNe light curves for 500 DECam FoVs for each follow-up.
The background population of simulated light curves consists of both type Ia and CC SNe that reach peak brightness in a range of MJDs extending from 30 nights before the observing window to 100 nights after the observing window has concluded. 
For the simulated type Ia SNe, we use templates from \citep[]{guy2010salt2} and rates from \citep[]{Dilday_2008}, and for the simulated CC SNe we use templates from \citep[]{Kessler_2010} and rates from \citep[]{Bernstein_2012}.
We do not include asteroids or AGN in the background sample since those models are not available within \snana, though it has been shown that they can be effectively removed with additional selection criteria \citep[]{Bailey:2007}. 
The simulated signal population consists of only CC SNe with an explosion time set to be the date of the IceCube neutrino alert.
Since explosion times are not observed, CC SNe models are more tightly constrained near peak brightness than during the rising phases, and we therefore use the flux relative to its peak value as a proxy for when the explosion began.
Within SNANA, we define the explosion time to be the earliest MJD for which the flux in any band reaches 1\% of the peak flux in the rest-frame of the SN.
To construct the signal sample, CC SN light curves are scanned for the MJD exhibiting a rise in flux above the 1\% peak flux level, and then shifted in time so that the MJD aligns with the desired explosion time.
By setting the neutrino trigger to the date of the start of the explosion, we are only considering prompt neutrino emission in this analysis, rather than neutrino emission that may occur during later stages of the collapse. 
Both sets of simulations account for the measured point spread function, zero point, and sky noise in each exposure to yield light curve quality representative of the actual DECam follow-up observations.
The simulations do not account for uncertainty in the CC SNe luminosity function, which is currently approximated using a Gaussian distribution with a mean and standard deviation determined from observed SNe luminosity \citep[]{cc_lum}. 

The process of background removal consists of a series of selection criteria applied to the simulated light curves. We list the criteria here and discuss the motivation for each cut in the remainder of this section.
\begin{enumerate}
\item \textit{Quality Cut}. Light curves must have detections on two separate nights, irrespective of the band of the detection. If a light curve passes this cut, we refer to it as ``optically detectable''.
\item \textit{Rising Cut}. Light curves must not have a detection on the first post-template night following the trigger, or if there is a detection, there must be an increase in flux of at least one magnitude in at least one band over the first two post-template nights. 
\item \textit{Phase Cut}. The peak MJDs predicted by the type Ibc and type II templates fit to the light curve must be at least 6 nights and 16 nights after the trigger, respectively.
\item \textit{Classify Cut}. Light curves must be classified as a CC SN by our Random Forest Classifier.
\end{enumerate}

{\bf Quality.} 
The quality cut is designed to guarantee DECam detectability from photometric data quality.
For this cut and the rising cut, we define a ``detection" in a given band and epoch to mean the photometric data has the following three properties: (1)  
\code{DiffImg} finds the object and does not mark it as a bad subtraction; 
(2) The ML score is larger than 0.7, meaning observation was determined to have good image quality and is unlikely to be an artifact; and (3) the signal-to-noise ratio is larger than 10.
The quality cut removes events that are photometrically of too poor quality to claim association with the neutrino by assuring that the event is observable with DECam. 
In the DECam FoV, we expect $17.5 \pm 0.3$ SNe and observe 11 objects for IC170922A, and for IC171106A we expect $10.0 \pm 0.3$ SNe and observe 10 objects. The under-fluctuation for IC170922A is mediated by the application of further cuts. 


{\bf Rising.}
The rising cut is designed to select light curves of recently-exploded objects by removing SNe that reach peak brightness before the neutrino trigger. 
The effectiveness of the rising cut is shown in the left panel of Figure \ref{fig:selection_criteria}.
For the two follow-ups presented here, the first observing night was used to make template images.
Therefore, ``post-template`` nights refer to all nights after the first night.
In future follow-ups if it is not possible to observe the field right away, or if template images already exist, this criterion will need to be reformulated.
After the rising cut, we expect $13.1 \pm 0.3$ SNe in the DECam FoV for IC170922A and observe 10 objects, and for IC171106A we expect $7.3 \pm 0.3$ SNe and observe 8 objects.

{\bf Phase.}
The most differentiating features between the signal and background populations that pass the first level of cuts are the type of SN and where in time the SN exploded relative to the neutrino trigger, which we refer to as the {\it phase} of the light curve. 
To exploit these features, we fit all light curves using the \snana implementation of the Photometric SuperNova IDentification tool \citep[\psnid]{0004-637X-738-2-162}, which offers not only best-fit phase information, but also fit probabilities and $\chi^2$ values for the type Ia, type Ibc, and type II templates fit to each light curve (henceforth $\chi^2_{\textrm{Ia}}$, $\chi^2_{\textrm{Ibc}}$, and $\chi^2_{\textrm{II}}$). 
The phase cut is designed to remove light curves that exploded before the trigger and are still rising during the observing window. 
The cutoffs of 6 and 16 nights were empirically derived from analyzing simulations. The motivation for these values is displayed in the center panel of Figure \ref{fig:selection_criteria}, and the performance of the phase cut is also shown in the left panel of the same figure.
After imposing the phase cut, we expect $5.6 \pm 0.2$ SNe in the DECam FoV for IC170922A and observe 2 objects, and for IC171106A we expect $2.74 \pm 0.17$ SNe and observe 2 objects.

\begin{figure}
\includegraphics[width=\columnwidth]{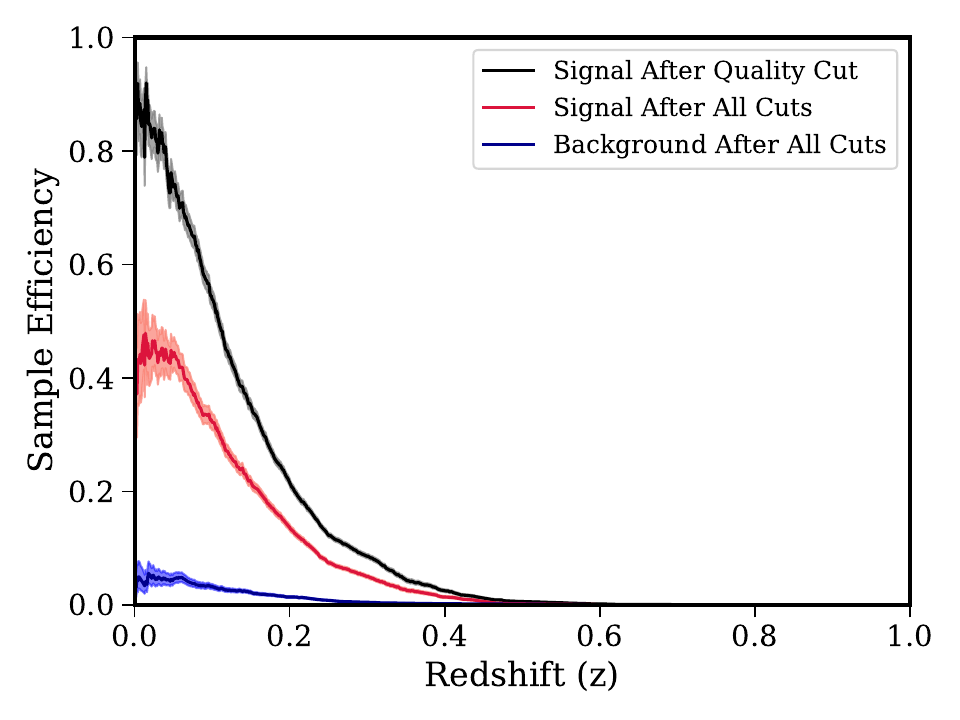}
\caption{Detection efficiency parameterized by redshift for the signal and background populations based on simulations of the IC170922A follow-up observations. The solid black line illustrates the fraction of CC SNe which would be detectable by DECam, while the solid red line shows the fraction of all CC SNe that pass our selection criteria. The shaded error regions give a 68\% confidence interval.
}
\label{fig:SensitivityPlots}
\end{figure}

{\bf Classify.}
Since \code{PSNID} was designed to make classifications with the goal of maximizing type Ia SNe purity, we opted to supplement the information used to make classifications with features that are tailored to the problem of selecting CC SNe.
Specifically, we found that by including features representative of the probability a light curve was a CC SNe, we were able to increase the fraction of known CC SNe classified as CC SNe (CC completeness) while simultaneously reducing the fraction of known type-Ia SNe classified as CC SNe (Ia false positive rate).
\code{PSNID} returns $\chi^2$ values for light curve fits to type-Ia $(\chi^2_{\textrm{Ia}})$, type-Ibc $(\chi^2_{\textrm{Ibc}})$, and type-II $(\chi^2_{\textrm{II}})$ templates, 
Based on this information, we define the normalized Ia and CC $\chi^2$ values $\bar{\chi}^2_{\textrm{Ia}}$ and $1 - \bar{\chi}^2_{\textrm{CC}}$, where 
\begin{align}\bar{\chi}^2_{\textrm{Ia}} &= \frac{\chi^2_{\textrm{Ia}}}{\chi^2_{\textrm{Ia}} + \chi^2_{\textrm{Ibc}} + \chi^2_{\textrm{II}}} \textrm{ \quad }\textrm{ and }\textrm{ } \\  \bar{\chi}^2_{\textrm{CC}} &= \frac{\textrm{min}(\chi^2_{\textrm{Ibc}}, \chi^2_{\textrm{II}})}{\chi^2_{\textrm{Ia}} + \chi^2_{\textrm{Ibc}} + \chi^2_{\textrm{II}}}.
\end{align}
Both of the features range from 0 to 1, and are expected to be closer to 0 for a type Ia SN while closer to 1 for a CC SN. 
Using these features, we reclassified the sample using a Random Forest Classifier \citep[]{Breiman2001} implemented with \code{Sci-Kit Learn} \citep[]{scikit-learn}. 
To construct the classifier, we first used Principal Component Analysis \citep[PCA]{Tipping:1999} with 8 components to project light curve fit information to the space of largest variance, and then employed an ensemble of 1000 decision tree classifiers with a restriction of 50 for the maximum tree depth.
The number of principal components, number of decision trees, and depth restriction on the decision trees were determined via an extensive search of the hyperparameter space using a separate validation dataset.

The classifier was trained using a \snana-simulated sample of CC light curves passing the quality, rising, and phase cuts that had an explosion time set to the neutrino trigger and a low redshift ($z<0.3$) as the target class. 
The background class was composed of CC and Ia light curves determined that passed the quality, rising, and phase cuts with no set restriction on the true explosion time of the SNe.
We also include spectroscopically-confirmed CC SNe and Ia SNe from the DES 3 Year SNe sample \citep[]{DESSN} in the training set signal and background samples, respectively.
The ratio of simulated SNe to real SNe in the training set was found to strongly correlate to classifier accuracy, and we determined the optimum ratio for each follow-up using independent validation sets.
We find that classification accuracy is optimized for both simulated and real SNe when 36\% (40\%) of the training set is simulated for IC170922A (IC171106A), and the determination of these ratios is shown in Figure \ref{fig:ev1_validation}.
The dependence in classification accuracy on the inclusion of real SNe is likely due to the fact that the real SNe were observed under better conditions than our follow-ups.
Since our classifier uses best-fit properties to make predictions, it is sensitive to how well the light curve fitting reflects the true properties of the explosion, so the better observing conditions of the real SNe could add new classification information that a simulated training set lacked.

In the process of training, we used stratified 5-fold cross validation to limit overfitting. 
This procedure trains on 4/5 of the training data, tests on the remaining 1/5, and then repeats the process so that each fifth of the training data is used once for testing; the purpose is to introduce small variations in the training set so that the classifier does not effectively memorize the training set.
The classifier operates at 83.7\% purity, 68.1\% completeness and with a 7.3\% false positive rate on the training set and at 81.8\% purity, 66.7\% completeness, and with a 8.1\% false positive rate on the testing set. 
The similarity of the performance of the classifier on familiar and unseen data is a good indicator that the classification is not suffering from overfitting and will be able to generalize to other datasets. 
A Receiver Operating Characteristic curve for our classifier is displayed in the right panel of Figure \ref{fig:selection_criteria} and shows the classifier operating with a mean area under curve (AUC) of $0.88$. The standard deviation of the AUC over the five folds of the training data is 0.02, indicating classifications are relatively insensitive to the representation of exact members of the training set.
After this final cut, we expect $0.74 \pm 0.07$ SNe in the DECam FoV for IC170922A and observe 1 object, and for IC171106A we expect $0.32 \pm 0.06$ SNe and observe 1 object.
These two remaining objects are our candidates from these first two follow-up efforts.

\begin{figure}
\includegraphics[width=\columnwidth]{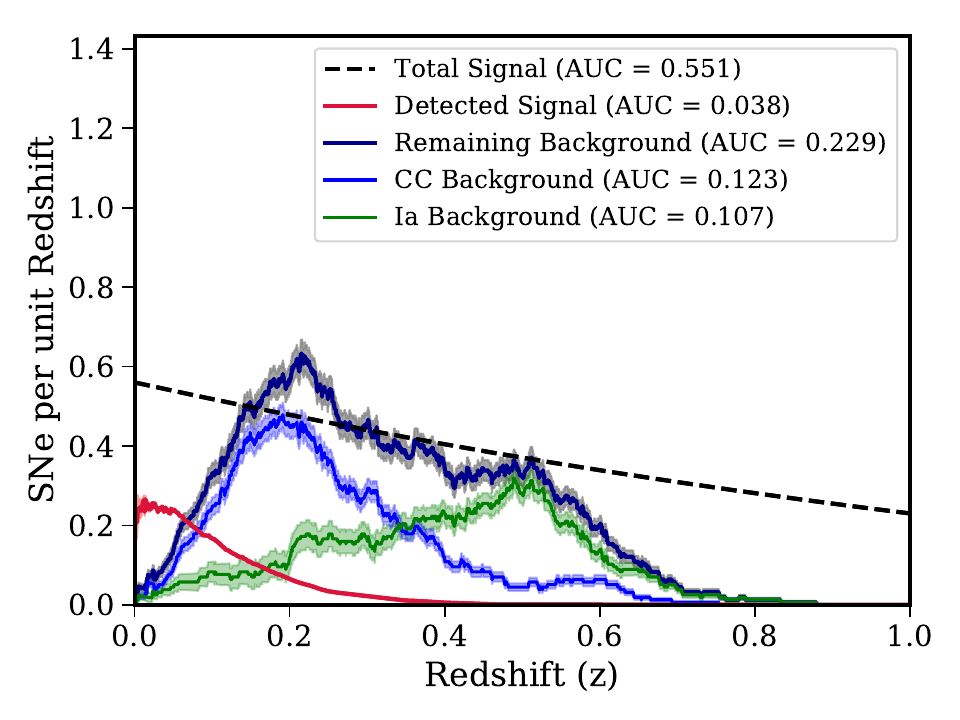}
\caption{Predicted number of events versus redshift based on simulations of the IC170922A follow-up after the selection criteria have been applied. Shaded error regions give a 68\% confidence interval for the signal curve and a standard Poisson statistical uncertainty for the background curves. The AUC shows the cumulative number of expected events in the solid angle of the IceCube 90\% containment region.}
\label{fig:ExpectedEvents}
\end{figure}

\subsection{Sensitivity Results}
We applied the selection criteria to our \snana simulations of the signal and background samples. For the signal samples we report the fraction of events passing each successive cut, as well as the fraction of optically-detectable events passing each successive cut. 
These fractions are multiplied by the CC SN rate to determine the expected number of signal-like events per IceCube alert as a function of redshift and scaled to the IceCube 90\% confidence level containment region (IC90). 
The number of background events passing each cut was normalized based on the number of FoVs generated to accurately reflect the expected number of background-like events per follow-up. These results are presented for each of the two events in Table \ref{tab:CutEfficiencies}.

\begin{figure}
\includegraphics[width=\columnwidth]{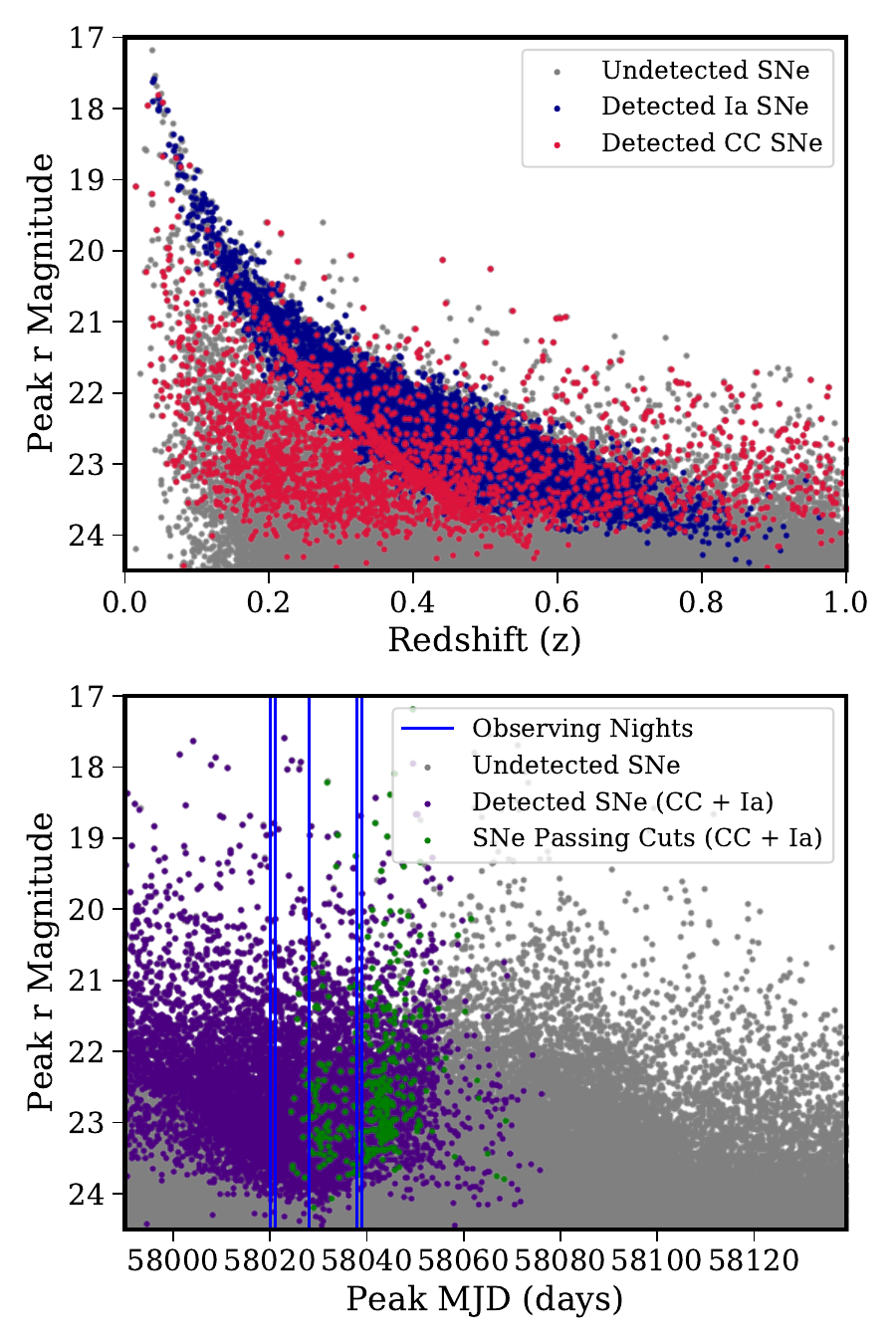}
\caption{The remaining SNe background events for the IC170922A simulations. \textit{Top}: Peak $r$-band magnitude versus the MJD of largest flux with reference MJD's of the IceCube alert and the DECam observing nights. \textit{Bottom}: Peak $r$-band magnitude versus redshift. Both figures display 500 times as many events as would be expected in a single DECam field of view.}
\label{fig:RemainingBG}
\end{figure}

For events with observing conditions of similar quality to the IC170922A follow-up, we expect to detect roughly 12.1\% of nearby ($z<0.3$), neutrino-emitting CC SNe using DECam and \texttt{NCIP}, while limiting the background to 0.23 unassociated SNe up to redshift $z=1$ within IC90. 
For the events similar to the IC171106A follow-up observations, we expect to detect 9.5\% of nearby signal events and 0.067 background events within IC90 per follow-up. 
The low signal detection efficiency is a result of the faintness of CC SNe and the magnitude limit of DECam, rather than the strictness of our selection criteria, evidenced by roughly 90\% of optically-detectable, low-redshift signal events passing all cuts across both events.
The remaining SNe background sample's magnitude, temporal, and redshift distributions are displayed in Figure \ref{fig:RemainingBG}.

Next, we consider the signal detection efficiency and number of remaining background events per set of follow-up observations as functions of redshift in order to calculate the maximum redshift to which we will be able to detect the potential CC SN counterpart of an IceCube alert. These relationships are shown for the IC170922A follow-up in Figure \ref{fig:SensitivityPlots}. 
Based on Figure \ref{fig:SensitivityPlots}, the fraction of signal sample events remaining after all cuts quickly diminishes as redshift increases, however, this behavior is largely a result of the low optical detectability of high redshift CC SNe.


The distribution of remaining background events per set of follow-up observations based on the observing conditions of the IC170922A follow-up is shown as a function of redshift Figure \ref{fig:ExpectedEvents}. 
We note that the type Ia SNe background has been suppressed in the redshift range of highest sensitivity (approximately $z < 0.2$). 
Therefore, only events that are similar to the signal population in both SN type and phase remain in the final SNe background population. 
The background sample of this analysis was normalized within \snana such that the number of events generated was equal to the expected number of SNe within a DECam FoV. The signal sample is normalized to one CC SNe, since only 1 or 0 signal CC SNe is possible per follow-up. The redshift distribution is derived by folding the detection efficiency as a function of redshift with the cosmic star formation rate as described in \secref{model}. We also multiply the signal distribution by the IceCube signalness value, which is the probability that the neutrino has an astrophysical origin.

From the signal and background redshift distributions, we integrate over all redshifts to obtain the total number of expected signal and background events for a given follow-up. 
These numbers are displayed on Figure \ref{fig:ExpectedEvents} as the Area Under Curve (AUC) values and become important when redshift information for pipeline candidates is unavailable, requiring us to account for all events along the line of sight. 
Next, we compare the expected number of detected CC SNe associated with the IceCube alert to the number of unassociated SNe expected to pass our selection criteria in a single follow-up observation. 
Figure \ref{fig:ExpectedEvents} displays this comparison for the IC170922A follow-up. 
The signal clearly dominates the background at low redshifts, and the cumulative number of signal events dominates the cumulative number of background events until roughly $z=0.1$.
Beyond this redshift, uncertainties in the modelling of SNe rise-times weaken the effect of the timing-based cuts (rising and phase) on reducing the background.
Additionally, as the redshift increases, CC SNe exploding at the time of the neutrino alert quickly become too faint to detect.
Together, both of these effects lead to background domination at high redshifts.

\section{Results of DECam Observations}
\label{sec:data_results}
We apply our candidate selection pipeline to the difference imaging results for the IC170922A and IC171106A follow-up efforts. 
We apply two additional selection criteria to our observations for effects that were not simulated. First we exclude candidates that move 0.1 arcseconds or more between observations to exclude asteroids. 
This cut did not remove anything that was not already removed by the quality cut. 
Second we perform a catalog search of all pipeline candidates and exclude objects with an angular separation less than 0.4 arcseconds from a known AGN or quasar. 
This criterion removed 1 candidate from each follow-up.
For each DECam follow-up, we display the number of potential candidates remaining after each successive cut in the rightmost column of Table \ref{tab:CutEfficiencies}. 
One difference imaging candidate passed all cuts from IC170922A (DES-Cand-1), and one candidate passed all cuts for IC171106A (DES-Cand-2). 

For the candidates found by \texttt{NCIP}, the coordinates and IceCube 90\% confidence intervals are shown in Figure \ref{fig:Localization}, the light curves are shown in \ref{fig:LightCurve}, and image stamps are shown in Figures \ref{fig:5430381_stamps_plots} and \ref{fig:332416_stamps_plots}.
As shown in Table \ref{tab:CutEfficiencies}, the number of expected SNe to pass our selection criteria per follow-up is between 0 and 1, so it is likely that these candidates are SNe. 
The light curves display a clear rise and the image cutouts show a transient object within a host galaxy.
A photometric redshift estimate was found for DES-Cand-2 using the methods of \citep{palmese} which employed a machine learning approach described in \citep{Sadeh_2016}.
This approach requires exposures in at least four bands, so we were not able to apply it to DES-Cand-1.

\begin{table}
\centering
\begin{tabular}{|c|c|c|c|c|}
\hline \hline
\multicolumn{5}{c}{\textit{IC170922A Follow-Up Candidates}} \\
\hline \hline
Name & RA (deg) & Dec (deg) & z & Prob. \\
\hline
DES-Cand-1 & 77.9885 & 6.7475 & -- & 0.0 \\ 
\hline
\multicolumn{3}{l}{Test Statistic:} & \multicolumn{2}{c}{0.0} \\
\multicolumn{3}{l}{p-value:} & \multicolumn{2}{c}{$>$ 0.999} \\
\hline \hline
\multicolumn{5}{c}{\textit{IC171106A Follow-Up Candidates}} \\
\hline \hline
Name & RA (deg) & Dec (deg) & z & Prob. \\
\hline
DES-Cand-2 & 340.3947 & 7.6820 & 0.31 & 0.1055 \\
\hline
\multicolumn{3}{l}{Test Statistic:} & \multicolumn{2}{c}{0.0558} \\
\multicolumn{3}{l}{p-value:} & \multicolumn{2}{c}{0.0250} \\
\hline \hline
\multicolumn{3}{l}{Global Test Statistic:} & \multicolumn{2}{c}{0.0143} \\
\multicolumn{3}{l}{Global p-value:} & \multicolumn{2}{c}{0.0930} \\
\hline \hline
\end{tabular}
\caption{\texttt{NCIP} candidates for each follow-up and results of maximum likelihood analysis. The uncertainties on the RA and Dec from \code{DiffImg} are negligible and the uncertainty on the redshift of DES-Cand-2 is $\pm0.02$. 
The rightmost column is the probability that the candidate is associated with the neutrino based on the expected signal and background events in the spatial bin of the candidate. The details of the likelihood analysis are presented in Appendix \ref{app:formalism}.
The p-values refer to the hypothesis test of CC SNe contributing the to IceCube neutrino flux as a population, rather than the association between the neutrino and candidate.}
\label{tab:candidates}
\end{table}

For the candidates selected by \texttt{NCIP}, we determined the probability that each candidate was associated with the neutrino source based on the numbers of expected signal and background events in the spatial bins of the candidate. 
This process is detailed in Appendix \ref{app:formalism}.
We also used the candidates to test the hypothesis that CC SNe contribute to the total IceCube neutrino flux. The significance levels of the hypothesis tests and the individual candidate probabilities are displayed in Table \ref{tab:candidates}.
Individually, the detected candidates are consistent with the expected backgrounds for their respective alerts derived in this analysis, hence we do not claim association between either of our candidates and their corresponding neutrino alerts.
Considering both follow-ups together, the observed number of SNe in the vicinity of the IceCube neutrino alert directions is compatible within $1\sigma$ to arise from fluctuations from the joint expected background.

\section{Follow-up Strategy Implications}
\label{sec:discussion}
In this section, we examine whether observational follow-up campaigns can be used to determine the CC SNe contribution to the TeV-PeV IceCube neutrino flux. 
We base the discussion on results and analysis
methods presented above and outline the requirements for a successful follow-up campaign.
We then expand on the analysis by forecasting the requisite number of follow-ups to make a statistically significant statement about this problem.


\subsection{Necessary Observational Components}

Previous efforts have been made to associate CC SNe with neutrino telescope alerts. 
ROTSE and TAROT were used in optical follow-ups of 42 ANTARES neutrino alerts, reaching a maximum $r$-band limiting magnitude of 18.6 with a field of view diameter of 2 deg \citep[]{1475-7516-2016-02-062}. 
One neutrino doublet IceCube alert has been followed up with ROTSE and PTF, and while a CC SN was found in the field of view, it was determined to be old and type IIn via spectroscopy from Keck I LRIS and Gemini/GMOS-N, and therefore unassociated with the neutrinos \citep[]{0004-637X-811-1-52}. 
The optical instruments in this effort reached a peak $r$-band limiting magnitude of 19.5 with a field of view diameter of 2.0 deg. 
Only a small fraction of cosmic neutrinos will be detected as doublets, while the large majority will be singlets. 
Accordingly, the follow-up of singlets is a promising approach.
However, as shown by Figure \ref{fig:RemainingBG}, significantly deeper observations are required in the singlet case.
Specifically, our \snana simulations, with peak $r$-band magnitudes displayed in Figure \ref{fig:RemainingBG}, show the importance of DECam’s imaging depth in searches for CC SNe since the fainter SNe in the 21-23.5 magnitude range become detectable. 
A complementary version of Figure \ref{fig:RemainingBG} where the limiting magnitude was set to $21 \textrm{ mag}$ is available in Figure \ref{fig:m21_bg}, illustrating the small fraction of all occurring SNe detectable at this optical depth.
Because CC SNe are expected to follow the cosmic massive star formation rate, they are expected to be mostly distributed at higher redshifts, and hence faint in the optical bands.
Therefore, it is not surprising that deep imaging offers a significant advantage to follow-up campaigns.

Even with the deep DECam observations presented in this work, detecting a CC SN as a TeV-PeV neutrino counterpart is still difficult as our sensitivity analysis suggests.
Our simulations and \texttt{NCIP} analysis show that only $\sim 9.5-12$ \% of associated CC SNe with $z\leq0.3$ would be bright enough to be detectable. 
For $z>0.3$, less than one percent of associated CC SNe can be detected.
When we fold this low detection efficiency with the cosmic massive star formation rate as described in \secref{model}, unassociated SNe are expected to overwhelm the signal population.
For example, based on the expected signal (0.038) and background (0.229) candidates per follow-up from Figure \ref{fig:ExpectedEvents} for IC170922A, we would detect approximately 4 associated CC SNe and 23 unassociated SN out to a redshift of $z=1.0$ within the IC90 regions over the course of 100 IceCube follow-up experiments using DECam.

The redshift distribution and SN type composition of the expected background after our candidate selection pipeline is displayed in Figure \ref{fig:ExpectedEvents}, and is particularly useful in understanding the difficulty of claiming association between a neutrino and a CC SNe. 
In Figure \ref{fig:ExpectedEvents}, the area under the dashed black curve represents the total expected
number of signal events for a given follow-up.

We have multiplied the redshift probability distribution function described by the cosmic massive star formation rate by the IceCube signalness for the alert, which was $\sim0.56$ for IC170922A. 
This curve also follows the assumption that the entire IceCube TeV-PeV energy neutrino flux is caused by CC SNe, which we refer to as the $\lambda=1.0$ case in Appendix \ref{app:formalism}.
We note that for low redshifts ($z \lesssim 0.2$), the redshifted neutrino energy will be very close to its rest frame energy. 
Therefore, we find it sufficient to leave out the contribution of the spectral energy distribution of neutrinos from the calculation. 
The red curve then folds the black curve with the signal detection efficiency from Figure \ref{fig:SensitivityPlots} to obtain the expected number of detected CC SNe TeV-PeV neutrino counterparts per follow-up as a function of redshift.
In the redshift range where the red curve is non-vanishing, we note that the most dominant background is CC SNe that are present in the FoV but are not associated with the neutrino.
These CC SNe have also passed the rising and phase cuts of our pipeline, meaning they appear to be temporally coincident with a neutrino as well.
Therefore, the largest background in this analysis is CC SNe that are temporally coincident with the neutrino, which unfortunately are the exact characteristics of our signal population.

Given this background, our main tool for determining if a given candidate is associated with the neutrino alert is the proximity of the candidate to the alert centroid. Additional differentiating features between signal and background samples are the candidate redshift and spectroscopic characterization. The redshift can likely be determined to sufficient accuracy based on a photometric redshift of the host galaxy, but if spectra can be obtained then it would also be possible to exclude the type Ia SNe background.

\begin{figure*}
\includegraphics[width=\textwidth]{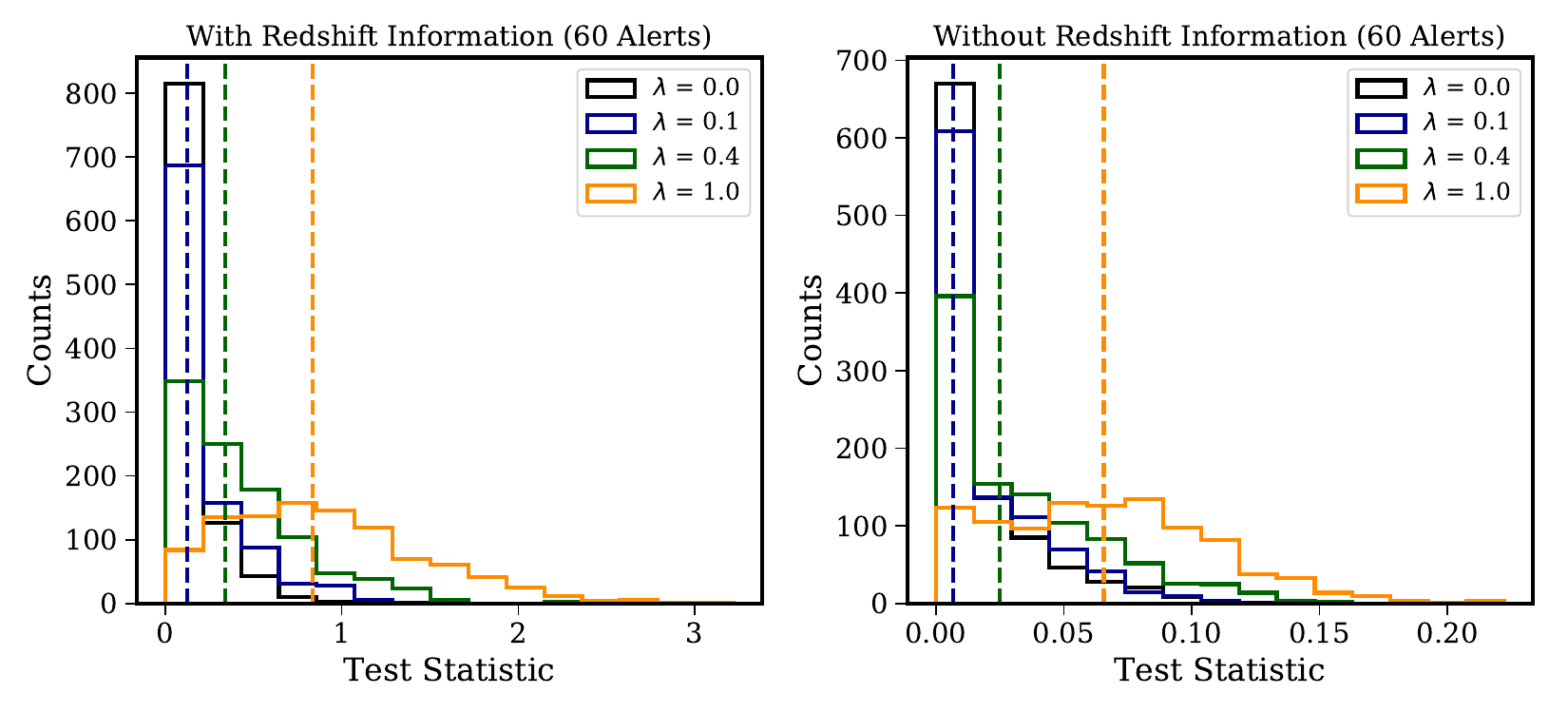}
\caption{Test statistic distributions for 1000 realizations of the follow-up of 60 IceCube alerts with DECam parameterized by the expected fraction of IceCube events caused by CC SNe ($\lambda$). 
The left/right panel displays the test statistic distribution with/without redshift information available for all alerts. The median values of the signal-hypothesis test statistic distributions are also displayed.}
\label{fig:TS_distribution}
\end{figure*}


\subsection{Forecasting Statistical Significance}
An individual follow-up with deep optical imaging, triggered observations, and redshift information for all candidates is still unlikely to reach a meaningful level of significance due to a signal-like background of unassociated CC SNe that are coincident in time with neutrino alerts and a low signal detection efficiency.
Therefore, we frame this analysis as seeking to detect an excess of CC SNe located near IceCube neutrino alerts in space and time over the course of multiple follow-ups.
We present a full calculation later in this section, but we outline the basis of it here and restrict our focus to the IC90 region for clarity; the full calculation utilizes the entire DECam Fov.
In the IC90 region, from Figure \ref{fig:ExpectedEvents} we expect 0.038 signal events and 0.229 background events for a follow-up similar to IC170922A and from Figure \ref{fig:ev2_events_pdf} we expect 0.047 signal events and 0.067 background events for a follow-up similar to IC171106A.
In the full calculation, if redshift information is available, the contribution to the total significance is weighted by the expected number of signal and background events at the particular redshift.
Overall, this process just reduces the contribution of background events to the total significance, so we can account for the effect of having redshift information for a candidate by dividing the expected background by a factor of $\sim3$ to simulate a smaller redshift range.
Applying this factor and averaging the results leads to a mean expected 0.043 signal events and a mean expected 0.049 background events.
We then estimate the significance after $N$ follow-ups using a Poisson distribution:
\begin{align}
\textrm{p-value } = 1 - \textrm{PoissonCDF(Observed} \vert \textrm{Expected)}.
\end{align}
Setting the ``Expected'' number of events to be $N$ $\times$ 0.049 for the background-only null-hypothesis and setting the ``Observed'' number of events to be $N$ $\times$ (0.043 + 0.049) enables us to evaluate the significance as a function of the number of follow-ups.
As well, one can see that increasing the number of follow-ups will increase the global significance.
In this brief calculation, for 100 follow-ups we would expect to reach a p-value of $\sim0.029$, which is between $2\sigma$ and $3\sigma$.
The addition of positional information and a more accurate accounting of redshift information in the full calculation both increase the significance beyond this simplified example.

To make this calculation more rigorous, we simulate pipeline candidate events with frequencies matching the mean expected event rates from the two follow-ups.
To determine the probability of a \texttt{NCIP} candidate being the source of the neutrino, we evaluate the candidate's position and redshift in the context of the expected signal and background population distributions of these quantities.
For the signal sample, we fit independent 2D asymmetric Gaussian distributions to the 90\% confidence level error bounds on the RA and Dec. We multiply our detection efficiency by the fraction of events that would fall on a DECam CCD to account for chip gaps and the few events that fall outside the DECam FoV.
For the background distribution, we adopt a uniform distribution of events on the individual DECam CCDs.
The redshift distributions for the signal and background samples are taken from Figure \ref{fig:ExpectedEvents}.
Using these distributions, we assess whether CC SNe contribute to the TeV-PeV neutrino flux by searching for an excess of temporally coincident CC SNe in our observations compared to the expected background contamination rate.
Appendix \ref{app:formalism} explains this process in greater detail and provides and overview of the maximum likelihood formalism used to quantify the likelihood of association between our pipeline candidates and IceCube neutrino alerts.

Using the maximum likelihood framework and the angular and redshift distributions for the signal and background samples, we perform 1000 realizations of optical follow-up campaigns of IceCube alerts. 
For each alert follow-up, we determine a list of candidates by sampling the signal and background event distributions.
The list of candidates has a probability of containing a signal event of 0.038 $\times$ 0.89 = 0.0338 for the IC170922A follow-up, where 0.038 is the expected number of detected signal events per follow-up from Figure \ref{fig:ExpectedEvents} and 0.89 is the fraction of the signal that falls on the DECam CCDs.
The number of expected background events per follow-up (0.717) is determined from the background AUC in Figure \ref{fig:ExpectedEvents} by scaling the value to the full DECam field of view. 
The number of background events to simulate is then determined by a Poisson probability distribution with a mean of 0.717.
Once the number of candidates has been determined, we sample the angular and redshift distributions to obtain a realistic representation of the follow-up.
We also introduce a signal normalization parameter, $\lambda$, which represents the fraction of the TeV-PeV neutrino flux caused by CC SNe, and perform the realizations under different values of $\lambda$.

The test statistic distributions for 1000 realizations of follow-up campaigns of 60 alerts based on the  conditions of both alerts are displayed in Figure \ref{fig:TS_distribution}.
We determine the significance level at which we would be able to claim CC SNe contribute to the TeV-PeV IceCube flux by taking the median value of our alternative hypothesis ($\lambda \neq 0.0$) distributions and calculating the fraction of the null hypothesis distribution ($\lambda = 0.0$) greater than the median test statistic. 
This fraction can be interpreted as the frequentist p-value.
Figure \ref{fig:TS_distribution} shows that our sensitivity for detecting an excess of CC SNe temporally and spatially coincident with IceCube alerts is greater when redshift information is available and if CC SNe make up a large fraction of the TeV-PeV neutrino flux.

\begin{figure}
\includegraphics[width=\columnwidth]{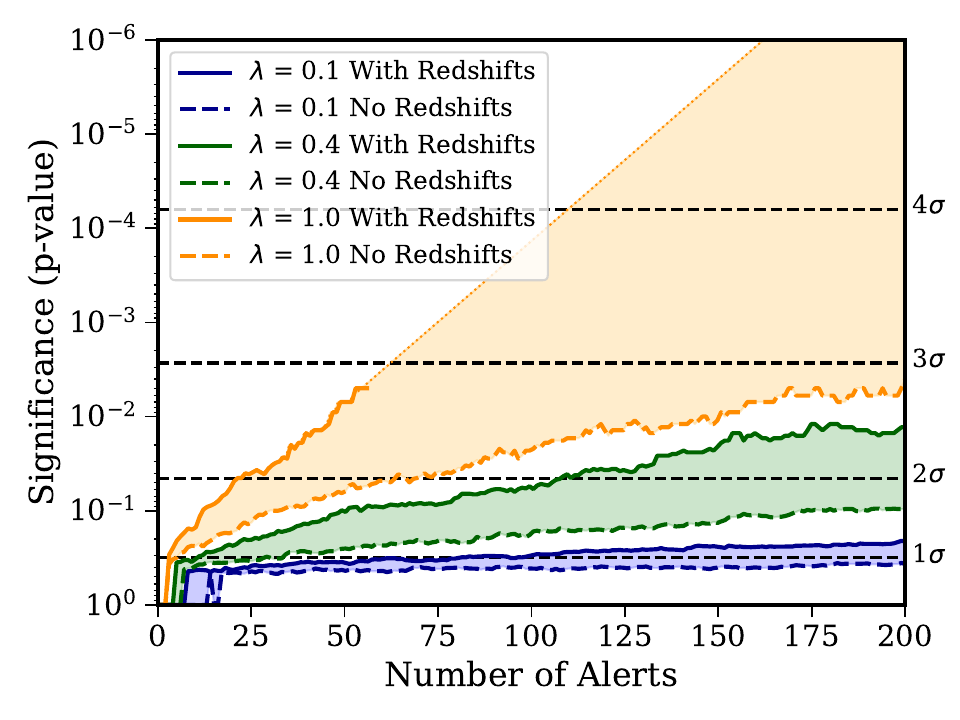}
\caption{Projected significance levels for DECam follow-up campaigns with increasing numbers of alerts based on the expected numbers of signal and background events in the IC170922A and IC171106A follow-ups. We display the relationship under different values of the parameter $\lambda$, which represents the true fraction of the IceCube TeV-PeV neutrino flux caused by CC SNe. We performed 1000 realizations of each follow-up campaign and calculate the p-value by comparing the median test statistic to the null hypothesis test statistic distribution. Since we only performed 1000 realizations, we are unable to probe p-values lower than 0.001, and therefore we extrapolate all p-values smaller than 0.001. }
\label{fig:ev1_significance}
\end{figure}

We then further the analysis of the viability of this type of study by calculating the discovery potential for follow-up campaigns with increasing numbers of alerts. 
The result of this calculation is displayed for the same three values of $\lambda$ in Figure \ref{fig:ev1_significance}. 
The cases of every candidate having available redshift information and no candidates having available redshift information are taken as the boundary cases of the best and worst case scenarios.
We note that a few alerts are required before the p-value is expected to differ from 1.0 in all cases, but that this behavior is explained by the fact that we use the median test statistic to calculate the p-value.
The p-value is fraction of null hypothesis test statistics strictly greater than the median of the signal hypothesis test statistic distributions, so the median must be larger than zero in order to produce a p-value less than one.
Due to the low detection efficiency of signal events, follow-up campaigns of few alerts are very likely to be statistically consistent with background and hence produce test statistic distributions that are predominantly composed of test statistics of zero.
Based on this calculation, optical follow-up campaigns of this nature would require approximately 60 follow-ups before an excess of temporally and spatially coincident CC SNe could be detected at the $3\sigma$ confidence level in the case that photometric redshift information is available for each candidate and that CC SNe account for 100\% of the TeV-PeV neutrino flux.
In the less optimistic case where CC SNe account for a smaller percentage of the TeV-PeV neutrinos, or redshift information is unable to be obtained, the required number of follow-ups can easily exceed 200.
The IceCube event rate is approximately 1 alert per month, and typically 1 in every three alerts is observable from CTIO based on solar and atmospheric conditions.

This analysis highlights several vital components for successful optical follow-ups of IceCube neutrino alerts searching for CC SNe.
First, deep imaging is needed to efficiently detect CC SNe counterparts to TeV-PeV neutrino at $z \sim 0.1$, as shown by the comparison between Figures \ref{fig:RemainingBG} and \ref{fig:ev2_SensitivityPlots}.
A sufficiently large FoV is also required such that the IceCube localization region can be covered.
DECam's $\sim23.5$ $r$-band limiting magnitude in 90 second exposures and $\sim3$~deg$^2$ FoV make it the current imager of choice in the southern hemisphere for this task.
Even with DECam, we anticipate a large background of CC SNe that are unassociated with the TeV-PeV neutrinos and difficult to distinguish from a neutrino-causing CC SNe.
Photometric redshifts and spectroscopic characterization can help with the differentiation, but with the low IceCube event rate and optical detection efficiency of CC SNe, a follow-up campaign would take several years to be able to observe a significant excess of CC SNe near Icecube alerts.
It is also possible that tightened constraints on the CC SNe luminosity function and SNe rise times will improve the accuracy of the simulations used in this analysis and in the significance forecasting.
It is anticipated that an expansion of the IceCube detector and/or improved realtime alert selection could increase the event rate and improve the angular resolution of neutrino reconstruction.

\section{Conclusion}
\label{sec:conclusion}
The real-time alert system for IceCube neutrinos has made possible the detection of electromagnetic counterparts for transient high-energy astrophysical events.
Using this system, we test whether CC SNe contribute to the largely unexplained TeV-PeV IceCube neutrino flux via a prompt relativistic jet mechanism during the collapse.
In this study, we followed-up two IceCube alerts from September 22, 2017 (IC170922A) and November 6, 2017 (IC170922A) with photometric observations in optical wavelengths using DECam.
These pilot observations combine the deepest-to-date optical follow-up observations of TeV-PeV IceCube neutrinos with a detailed estimation of the relevant backgrounds and a maximum likelihood analysis to fully characterize our sensitivity to CC SNe.

We developed and validated an automated candidate selection pipeline (\texttt{NCIP}) based on SNe simulations matched to our observing cadence and conditions.
Between the two alerts, 
\texttt{NCIP} found 2 plausible neutrino source candidate SNe (DES-Cand-1 for IC170922A and DES-Cand-2 for IC171106A) based on light curve properties and temporal coincidence. 
Both candidate SNe are located outside the respective 90\% confidence localization regions from IceCube.
Applying a maximum likelihood analysis to our candidates based on the simulated SNe background and spatial coincidence, we find that the two observed candidates are consistent with background unassociated SNe (p-value = 0.093).
Other multimessenger observations suggest that IC170922A is associated with the flaring gamma-ray blazar TXS 0506+056.

To assess the viability of continued optical searches triggered by IceCube alerts, we perform 1000 realizations of follow-up campaigns of up to 200 alerts.
We find that approximately 60 follow-ups are required to reach the $3\sigma$ confidence level in the case that the astrophysical neutrino flux is contributed entirely by CC SNe and redshifts are available for all detected candidates (\figref{ev1_significance}).
Without the availability of photometric redshift information for all candidates, the requisite number of follow-ups rises to $\sim200$ to make the same statement.
Furthermore, if neutrinos from relativistic jets inside CC SNe do not make up a large percentage of the TeV-PeV neutrino flux, the number of follow-ups required to achieve a high confidence detection could be much larger.
The sensitivity is limited primarily by the rate of unassociated low-redshift CC SNe that explode in spatial and temporal coincidence with the neutrino alerts. 
This residual background is degenerate with the signal population, and therefore challenging to further reduce.
Based on the methods presented here and the current rate of HESE and EHE realtime alerts from IceCube, a sustained optical follow-up program using a few hours of DECam time per semester extending over $\gtrsim10$ years would be needed to determine whether CC SNe contribute to the neutrino flux.

\section*{Acknowledgments}

R. Morgan thanks the LSSTC Data Science Fellowship Program, which is funded by LSSTC, NSF Cybertraining Grant \#1829740, the Brinson Foundation, and the Moore Foundation; his participation in the program has benefited this work.

All authors thank
Sahar	Allam,
Agne\`s	Fert\'e,
Indiarose	Friswell,
Kate	Furnell,
Mandeep	Gill,
Michael	Johnson,
Nikolay	Kuropatkin,
Marty	Murphy,
Douglas	Tucker,
Noah	Weaverdyck,
Reese	Wilkinson,
Hao-Yi	Wu, and
Brian	Yanny
for operating {\it Blanco/ DECam} and  for their help in performing the follow-ups presented in this work.
We also thank 
Douglas Finkbeiner, 
Abi Saha, 
Malhorta Sangeeta, 
Eddie Schlafly,  
Monika Soraisam, 
Scott Sheppard, 
and Catherine Zucker for operating {\it Blanco/ DECam} during more recent follow-ups.





This work was completed in part with resources provided by the University of Chicago Research Computing Center.

Funding for the DES Projects has been provided by the U.S. Department of Energy, the U.S. National Science Foundation, the Ministry of Science and Education of Spain, 
the Science and Technology Facilities Council of the United Kingdom, the Higher Education Funding Council for England, the National Center for Supercomputing 
Applications at the University of Illinois at Urbana-Champaign, the Kavli Institute of Cosmological Physics at the University of Chicago, 
the Center for Cosmology and Astro-Particle Physics at the Ohio State University,
the Mitchell Institute for Fundamental Physics and Astronomy at Texas A\&M University, Financiadora de Estudos e Projetos, 
Funda{\c c}{\~a}o Carlos Chagas Filho de Amparo {\`a} Pesquisa do Estado do Rio de Janeiro, Conselho Nacional de Desenvolvimento Cient{\'i}fico e Tecnol{\'o}gico and 
the Minist{\'e}rio da Ci{\^e}ncia, Tecnologia e Inova{\c c}{\~a}o, the Deutsche Forschungsgemeinschaft and the Collaborating Institutions in the Dark Energy Survey. 

The Collaborating Institutions are Argonne National Laboratory, the University of California at Santa Cruz, the University of Cambridge, Centro de Investigaciones Energ{\'e}ticas, 
Medioambientales y Tecnol{\'o}gicas-Madrid, the University of Chicago, University College London, the DES-Brazil Consortium, the University of Edinburgh, 
the Eidgen{\"o}ssische Technische Hochschule (ETH) Z{\"u}rich, 
Fermi National Accelerator Laboratory, the University of Illinois at Urbana-Champaign, the Institut de Ci{\`e}ncies de l'Espai (IEEC/CSIC), 
the Institut de F{\'i}sica d'Altes Energies, Lawrence Berkeley National Laboratory, the Ludwig-Maximilians Universit{\"a}t M{\"u}nchen and the associated Excellence Cluster Universe, 
the University of Michigan, the National Optical Astronomy Observatory, the University of Nottingham, The Ohio State University, the University of Pennsylvania, the University of Portsmouth, 
SLAC National Accelerator Laboratory, Stanford University, the University of Sussex, Texas A\&M University, and the OzDES Membership Consortium.

Based in part on observations at Cerro Tololo Inter-American Observatory, National Optical Astronomy Observatory, which is operated by the Association of Universities for Research in Astronomy (AURA) under a cooperative agreement with the National Science Foundation.

The DES data management system is supported by the National Science Foundation under Grant Numbers AST-1138766 and AST-1536171.
The DES participants from Spanish institutions are partially supported by MINECO under grants AYA2015-71825, ESP2015-88861, FPA2015-68048, SEV-2012-0234, SEV-2016-0597, and MDM-2015-0509, 
some of which include ERDF funds from the European Union. IFAE is partially funded by the CERCA program of the Generalitat de Catalunya.
Research leading to these results has received funding from the European Research
Council under the European Union's Seventh Framework Program (FP7/2007-2013) including ERC grant agreements 240672, 291329, and 306478.
We  acknowledge support from the Australian Research Council Centre of Excellence for All-sky Astrophysics (CAASTRO), through project number CE110001020.

This manuscript has been authored by Fermi Research Alliance, LLC under Contract No. DE-AC02-07CH11359 with the U.S. Department of Energy, Office of Science, Office of High Energy Physics. The United States Government retains and the publisher, by accepting the article for publication, acknowledges that the United States Government retains a non-exclusive, paid-up, irrevocable, world-wide license to publish or reproduce the published form of this manuscript, or allow others to do so, for United States Government purposes.

\facility{Blanco (DECam)}
\software{
\psnid \citep[]{0004-637X-738-2-162},
\snana \citep{Kessler:2009},
\code{matplotlib} \citep{Hunter:2007},
\code{pandas} \citep[]{pandas:2010},
\code{numpy} \citep{numpy:2011}, 
\code{sci-kit learn} \citep[]{scikit-learn},
\code{scipy} \citep{scipy:2001}
}

\clearpage

\appendix
\numberwithin{figure}{section}
\numberwithin{table}{section}

\section{Maximum Likelihood Formalism}
\label{app:formalism}
In this appendix we describe the formalism used to quantify the statistical power of our analysis in terms of testing the hypothesis that CC SNe contribute to the IceCube high-energy neutrino flux. We perform a maximum likelihood analysis in redshift-position space to determine the significance of observed excesses of CC SNe at the location and times of IceCube neutrino alerts. 

\begin{figure}
\centering
\begin{minipage}{.45\textwidth}
  \centering
  \includegraphics[width=\linewidth]{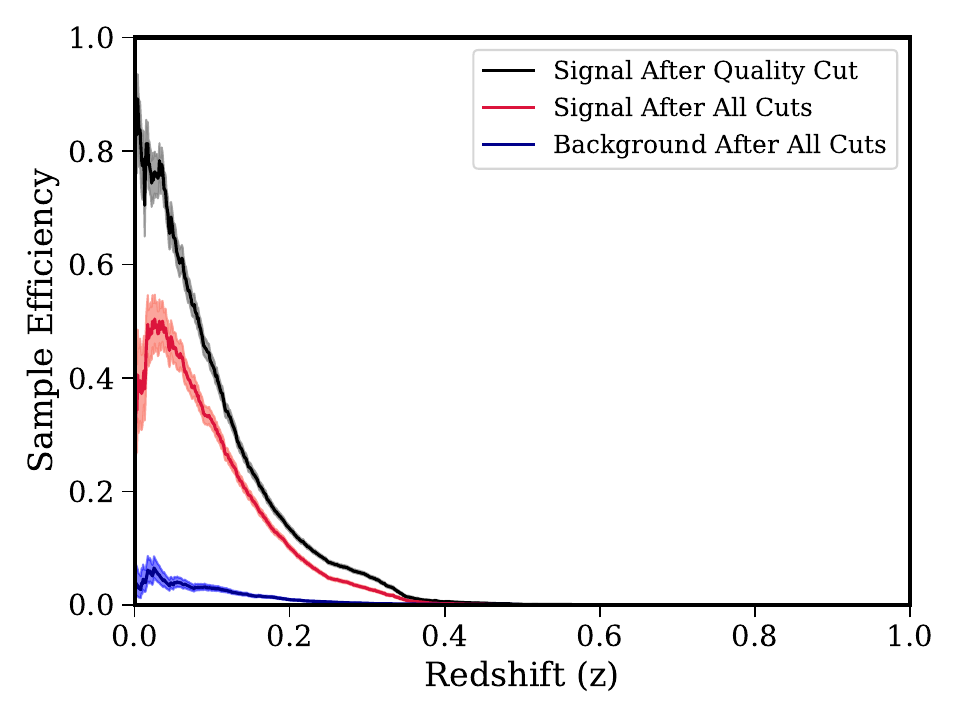}
  \caption{ Detection efficiency parameterized by redshift for the signal and background populations for the IC171106A alert follow-up. 
  }
  \label{fig:ev2_SensitivityPlots}
\end{minipage}%
\hspace{0.06\textwidth}
\begin{minipage}{.45\textwidth}
  \centering
  \includegraphics[width=\linewidth]{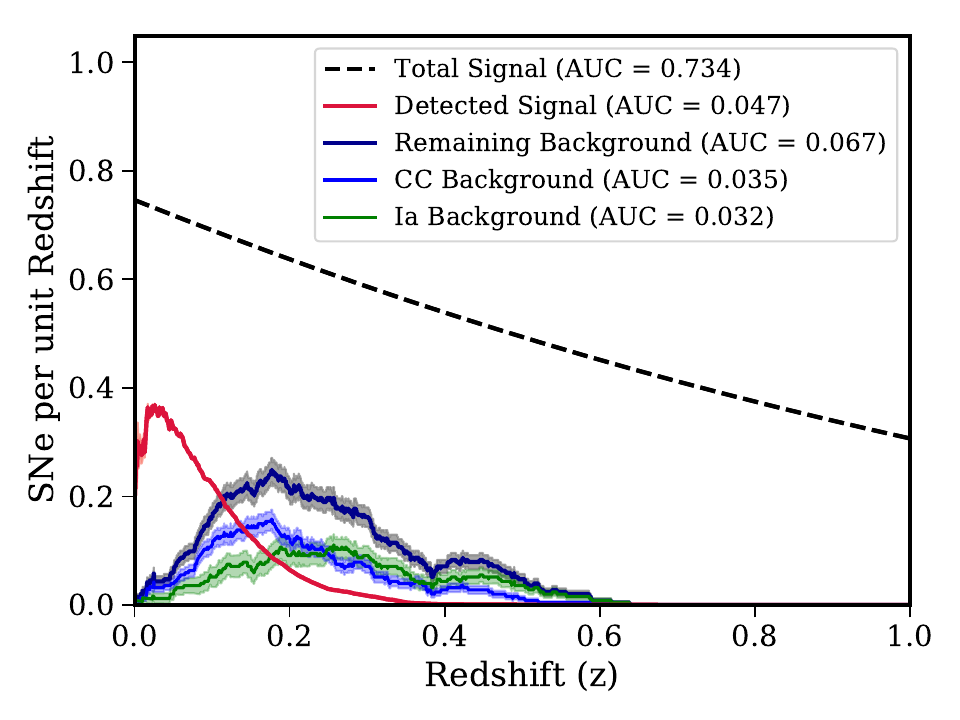}
  \caption{Expected events at a given redshift based on simulations of the IC171106A follow-up. \\}
  \label{fig:ev2_events_pdf}
\end{minipage}
\end{figure}

From our \code{SNANA} simulations of the two events, we have determined the redshift distribution of our expected signal and background populations. These distributions are displayed in Figure \ref{fig:ExpectedEvents} for IC170922A and in Figure \ref{fig:ev2_events_pdf} for IC171106A.
The black curves in these figures will be referred to as $\xi S_z(z)$, where $\xi$ is the IceCube signalness and $S_z(z)$ is the probability density function for signal events in redshift space given by \eqnref{redshift}.
The number of detected signal events per unit redshift (red curve in these figures) is derived from the black curve through a multiplication by the optical detection efficiency as a function of redshift $\varepsilon(z)$, so we express this quantity as $\xi S_z(z) \varepsilon(z)$.
It is necessary at this point to introduce a signal normalization parameter $\lambda$, which will represent the total fraction of the IceCube TeV-PeV flux caused by CC SNe. Therefore, over several follow-ups, the expectation value of $\lambda$ will be the fraction of IceCube neutrinos caused by CC SNe.
Thus, the expected number of signal events per unit redshift is $\lambda \xi S_z(z) \varepsilon(z)$.
The expected number of background events per unit redshift is represented by the curves labelled ``Remaining Background'' in Figures \ref{fig:ExpectedEvents} and \ref{fig:ev2_events_pdf}, and we express this function in our formalism as $B_z(z)$.

We distribute the events in position space according to a uniform distribution across the DECam field of view for the background population and according to a 2D asymmetric Gaussian fit to the 90\% confidence interval containment region determined by IceCube for the signal population.
The signal and background distributions for angular space are expressed as $S_\Omega(\Omega)$ and $B_\Omega(\Omega)$, which are normalized probability distribution functions. 
Therefore, $\lambda \xi S_z(z) \varepsilon(z) S_\Omega(\Omega)$ is the expected number of signal events per unit redshift per unit solid angle selected by our pipeline for a given follow-up, and $B_z(z) B_\Omega(\Omega)$ is the expected number of background events per unit redshift per unit solid angle selected by our pipeline for a given follow-up.

Suppose that after a single follow-up, we find one candidate (which we will denote as Candidate $i$) that passes all selection criteria. Consider two cases: redshift information being available for Candidate $i$ and redshift information not being available for Candidate $i$. In these two separate cases, we can write the model-expected numbers of events as
\vspace{-0.2cm}

\begin{table}[h]
\centering
\begin{tabular}{ll}
$m_i(\lambda) = \int dz d\Omega \textrm{ }  \delta(z-z_i)\delta(\Omega-\Omega_i)\left[\lambda \xi S_z(z) \varepsilon(z) S_\Omega(\Omega) + B_z(z) B_\Omega(\Omega)\right]$ &with redshift information and\\
$m_i(\lambda) = \int dz d\Omega \textrm{ }  \delta(\Omega-\Omega_i)\left[\lambda \xi S_z(z) \varepsilon(z) S_\Omega(\Omega) + B_z(z) B_\Omega(\Omega)\right]$ &without redshift information,
\end{tabular} 
\end{table}

\vspace{-0.4cm}
\noindent where we integrate from redshift $z=0$ to $z=1$ and over the entire DECam FoV. The delta functions serve to evaluate the expected numbers of signal and background event functions at the position and redshift of Candidate $i$. 
Therefore, in the case without available redshift information, we get contributions from all redshifts between $z=0$ and $z=1$.
To make the discussion more straightforward, we will adopt the following shorthand notation:
\begin{align*}
S_i &\equiv \int dz d\Omega \textrm{ }  \delta(z-z_i)\delta(\Omega-\Omega_i)\xi S_z(z) \varepsilon(z) S_\Omega(\Omega) \quad \textrm{ and}\\
B_i &\equiv \int dz d\Omega \textrm{ }  \delta(z-z_i)\delta(\Omega-\Omega_i)B_z(z) B_\Omega(\Omega) \textrm{,}
\end{align*}
where it is to be understood that in the case without available redshift information for Candidate $i$, $S_i$ and $B_i$ are defined without the $\delta(z-z_i)$ term, implying the need to include the contributions from all redshifts. Therefore, we can write $m_i(\lambda) = \lambda S_i + B_i$.

Now assuming the model of expected counts follows a Poisson distribution, we can write the likelihood and log-likelihood of observing $k$ candidates in a given follow-up as the product of probabilities of observing $k_j$ objects in redshift-spatial bin $j$ as
\begin{align*}
    \mathcal{L} =& \prod_{j=1}^{N \textrm{ bins}}\frac{[m_j(\lambda)]^{k_j} e^{-m_j(\lambda)}}{k_j !} \\
    \implies \log(\mathcal{L}) =& \sum_{j=1}^{N \textrm{ bins}} \log \left( \frac{[m_j(\lambda)]^{k_j} e^{-m_j(\lambda)}}{k_j !} \right) \\
    \implies \log(\mathcal{L}) =& \sum_{j=1}^{N \textrm{ bins}} \left( k_j \log (m_j(\lambda)) - m_j(\lambda) - \log(k_j !) \right) \textrm{.}
\end{align*}
In the limit of vanishingly small redshift-spatial bins where each bin contains either one or zero objects, we can express the log-likelihood as
\begin{align*}
    \log(\mathcal{L}) =& \left[\sum_{i=1}^{N \textrm{ candidates}}\log(m_i(\lambda))\right] - \int dz d\Omega \textrm{ } m(\lambda)\textrm{,} 
\end{align*}
where the first term comes from evaluating the $k_j\log(m_j(\lambda))$ and $\log(k_j!)$ terms for $k_j=0$ and $k_j=1$, and the second term is the evaluation of the sum of expected model counts overall all bins, regardless of the presence of candidates. 
Dropping all terms that are independent of $\lambda$, since they do not contribute in the maximization, we can write
$$ \log(\mathcal{L}) = \left[\sum_{i=1}^{N \textrm{ candidates}} \log(\lambda S_i + B_i)\right] - f\lambda, \quad \textrm{ where } \quad f \equiv \int dz d\Omega \textrm{ } \xi S_z(z) \varepsilon(z) S_\Omega(z).$$
The value of $f$ can be recognized as the area under the ``Detected Signal'' curve in Figures \ref{fig:ExpectedEvents} and \ref{fig:ev2_events_pdf} multiplied by the fraction of signal events that would fall on a CCD in our 2D asymmetric Gaussian distribution of signal events.

Maximizing the log likelihood with respect to $\lambda$ via $\partial \log(\mathcal{L}) / \partial \lambda = 0$ and multiplying by $\hat{\lambda}$, we obtain a maximization condition in terms of the individual probabilities for each candidate to be associated with the neutrino alert:
$$ f\hat{\lambda} = \sum_{i=1}^{N \textrm{ candidates}} \frac{\hat{\lambda} S_i}{\hat{\lambda} S_i + B_i} \equiv \sum_{i=1}^{N \textrm{ candidates}} p_i \textrm{,}$$
where $\hat{\lambda}$ is the optimum value of $\lambda$ that satisfies this above condition.
Operating under this condition that maximizes the log-likelihood, we can use the delta-log-likelihood to characterize the significance of a follow-up detecting a certain amount of excess temporally and spatially coincident CC SNe:
$$\Delta\log(\mathcal{L}) = \log(\mathcal{L})|_{\lambda = \hat{\lambda}} - \log(\mathcal{L})|_{\lambda = 0}\textrm{.}$$
Over a sequence of follow-up observations, then, we define our test statistic (TS) to be the sum of the delta-log-likelihoods of all the constituent follow-ups.
$$\textrm{ TS } = \sum_{n=1}^{N \textrm{ follow-ups}} \left(\Delta \log(\mathcal{L})\right)_n.$$

For a follow-up campaign that observes a given number of SNe candidates, we can evaluate the significance of the claim that the number of candidates observed is greater than the expected background and hence likely linked to the IceCube neutrino alerts by comparing the calculated TS value to the null-hypothesis TS distribution. 
In this work, we obtain the null-hypothesis TS distribution by performing 1000 realizations of follow-up campaigns with $\lambda$ set to 0.0. 
To obtain our signal-hypothesis TS value for a given value of $\lambda$ we perform the 1000 realizations and take the median of the distribution to be our signal TS value. 
The p-value for this test is the fraction of the null-hypothesis distribution that is larger than the median signal-hypothesis TS value, allowing for a direct evaluation of the significance level at which we can claim CC SNe contribute to the TeV-PeV IceCube neutrino flux.

\newpage

\section{Additional figures}
\label{app:example}

\begin{figure}[h]
\centering
\includegraphics[width=0.6\textwidth,trim={2cm 3cm 2cm 3cm},clip]{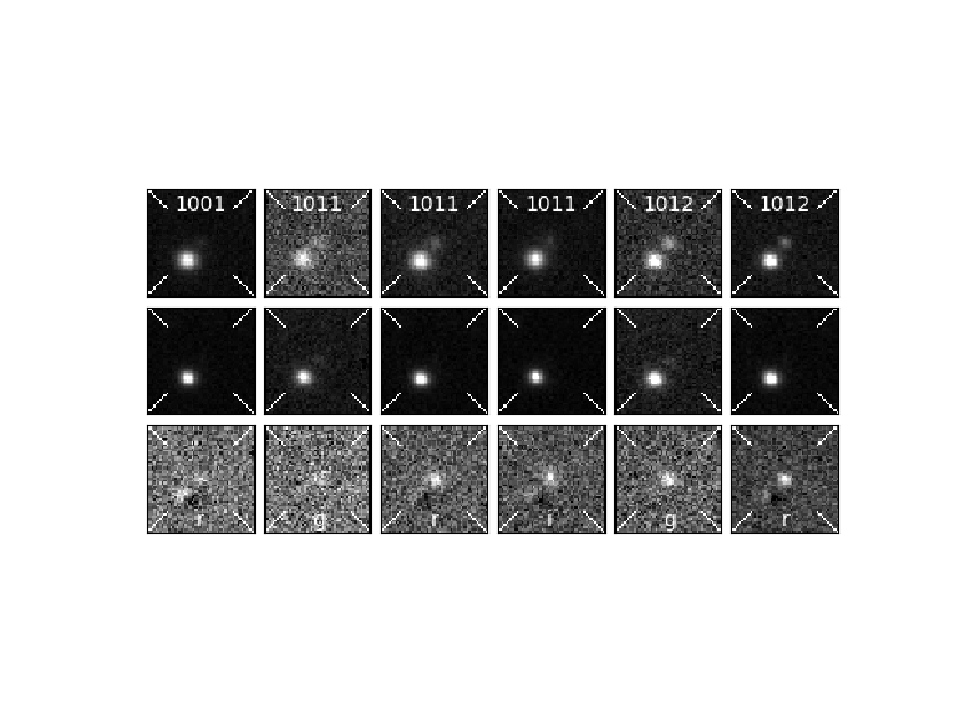}
\caption{Image stamps for DES-Cand-1. The top row is the search image, the middle row is the template, and the bottom row is the difference image. Filters are shown at the bottom of each column and dates are given in MMDD format at the top of each column.}
\label{fig:5430381_stamps_plots}
\end{figure}



\begin{figure}
\centering
\includegraphics[width=0.6\textwidth,trim={2cm 2.5cm 2cm 2.5cm},clip]{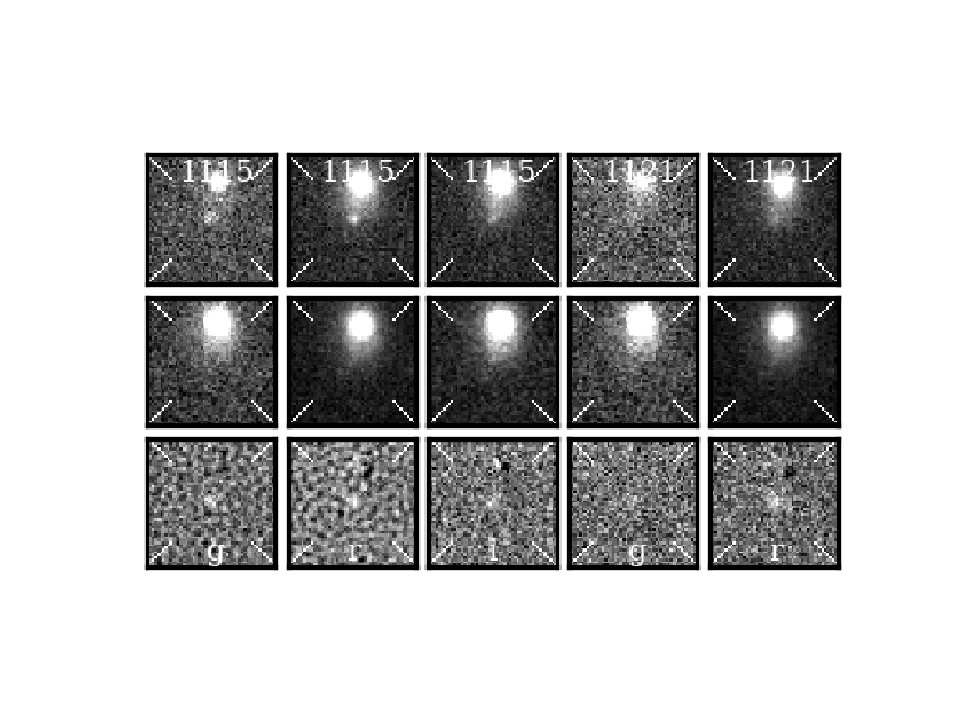}
\caption{Image stamps for DES-Cand-2. The top row is the search image, the middle row is the template, and the bottom row is the difference image. Filters are shown at the bottom of each column and dates are given in MMDD format at the top of each column.}
\label{fig:332416_stamps_plots}
\end{figure}


\begin{figure}
\centering
\includegraphics[width=0.9\textwidth]{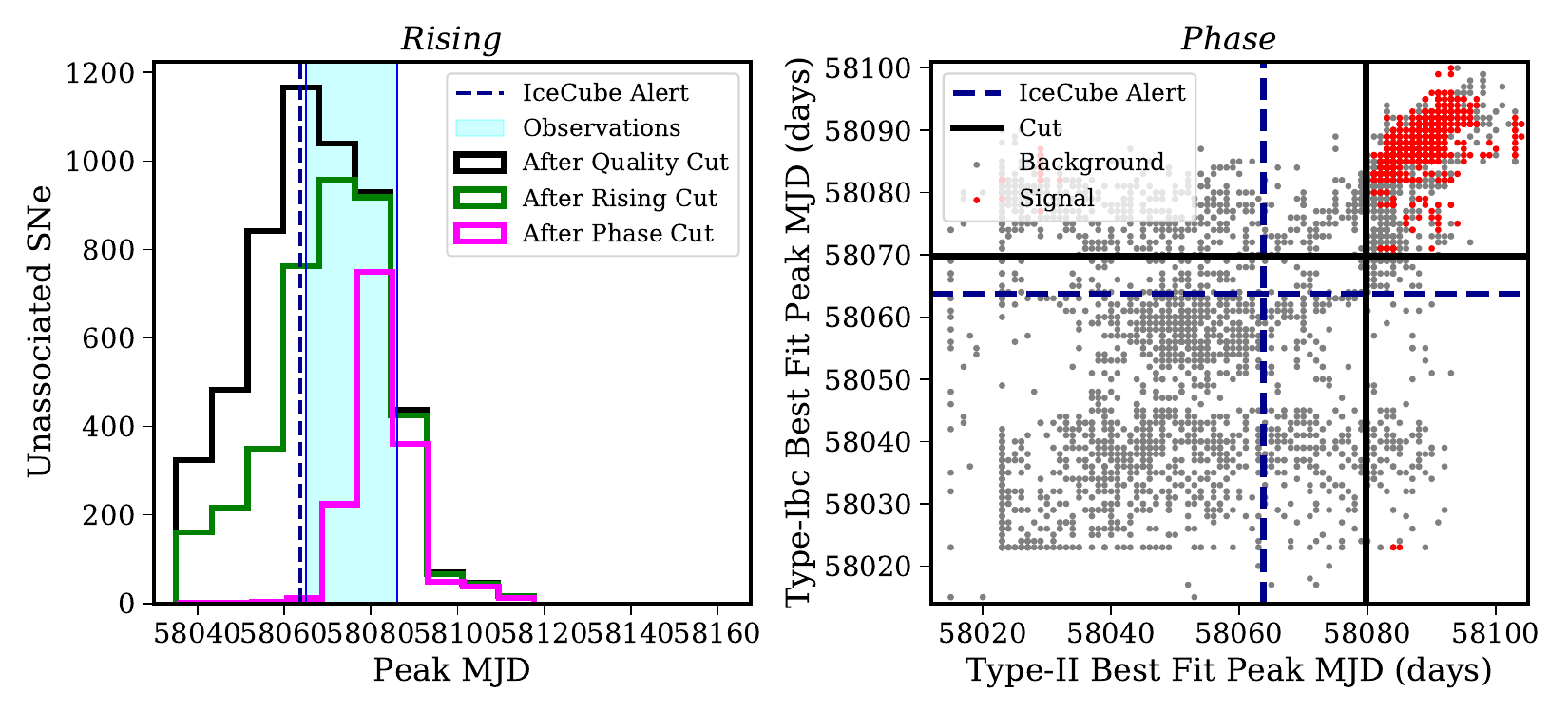}
\caption{The efficacies of the rising and phase cuts applied to the IC171106A simulations. The number of reported events is amplified by a factor of 500 as compared to a typical DECam field of view.}
\label{fig:ev2_cuts_plots}
\end{figure}

\begin{figure}
\centering
\begin{minipage}{.45\textwidth}
  \centering
  \includegraphics[width=\linewidth]{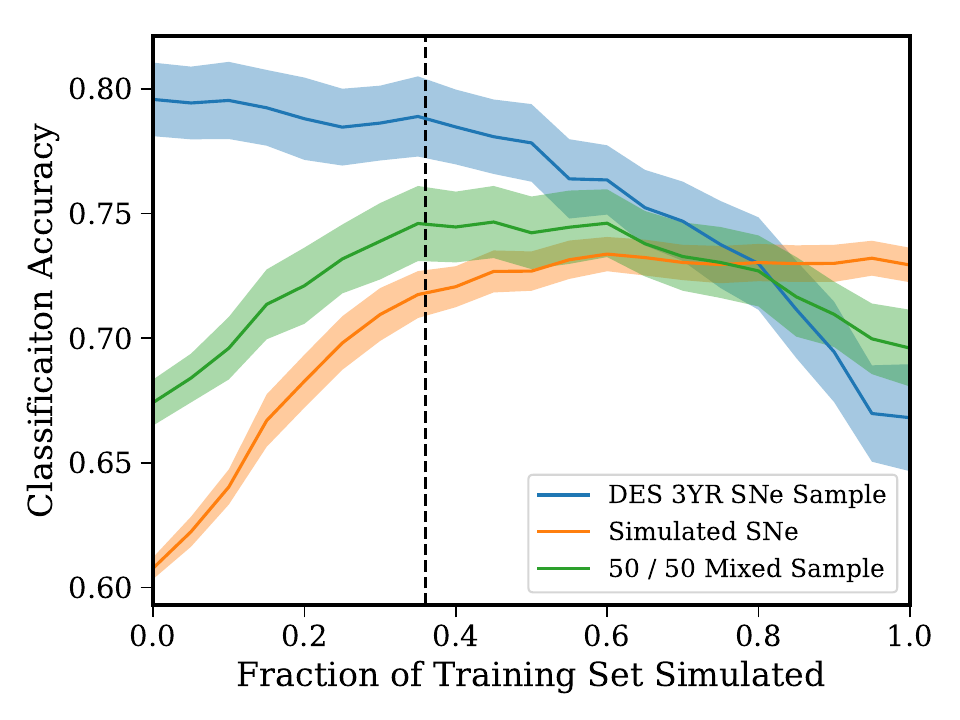}
  \label{fig:ev1_validation}
\end{minipage}
\hspace{0.06\textwidth}
\begin{minipage}{.45\textwidth}
  \centering
  \includegraphics[width=\linewidth]{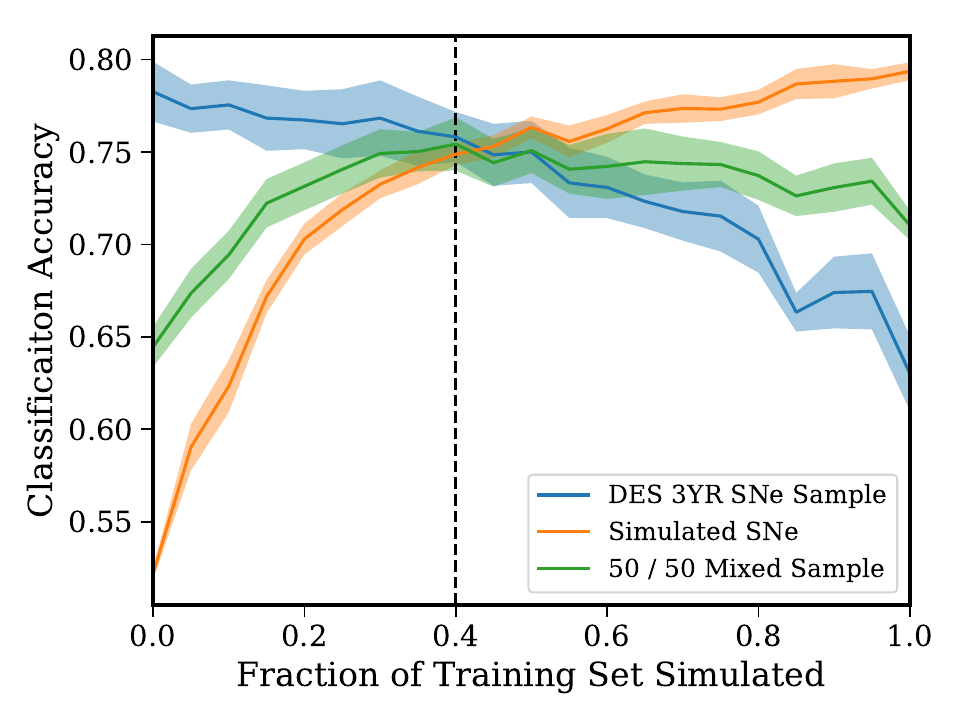}
  \label{fig:ev2_validation}
\end{minipage}%
\caption{Classifier accuracy for real SNe, simulated SNe, and a mixed sample parameterized by the fraction of the training set composed of simulated SNe. The optimum fraction chosen for training is shown by the black dashed line. \textit{Left:} Based on the IC170922A follow-up. \textit{Right:} Based on the IC171106A follow-up.}
\end{figure}

\begin{figure}
\centering
\begin{minipage}{.45\textwidth}
  \centering
  \includegraphics[width=\linewidth]{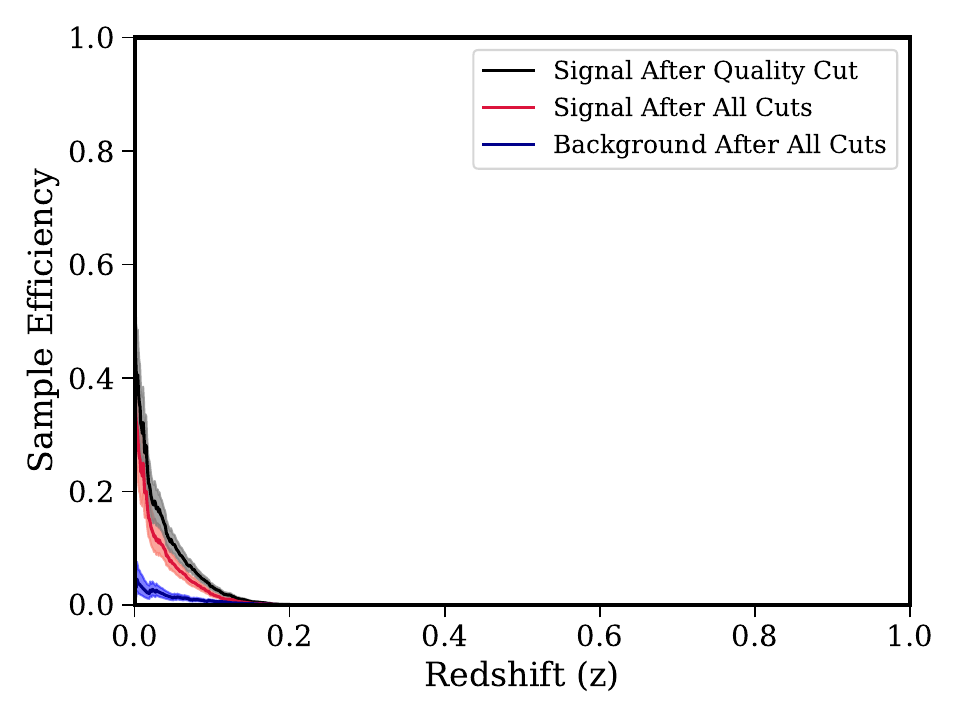}
  \caption{The detection efficiency as a function of redshift for simulations of IC170922A with a limiting magnitude of $21 \textrm{ mag}$. \\ \\ }
  \label{fig:m21_sens_results}
\end{minipage}
\hspace{0.06\textwidth}
\begin{minipage}{.45\textwidth}
  \centering
  \includegraphics[width=\linewidth]{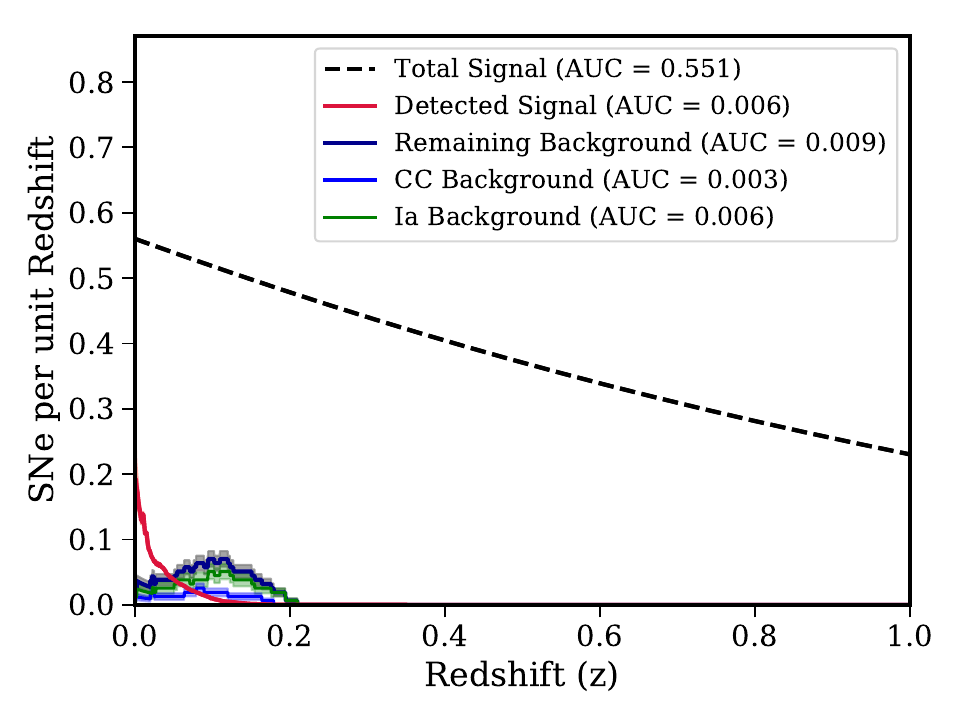}
  \caption{Expected events at a given redshift based on simulations of the IC170922A follow-up with limiting magnitue 21. Shaded error regions give a 68\% confidence interval for the signal curve and a standard Poisson statistical uncertainty for the background curves.}
  \label{fig:ev2_events_pdf.png}
\end{minipage}%
\end{figure}

\begin{figure}
\centering
\begin{minipage}{0.45\textwidth}
  \centering
  \includegraphics[width=1.02 \linewidth]{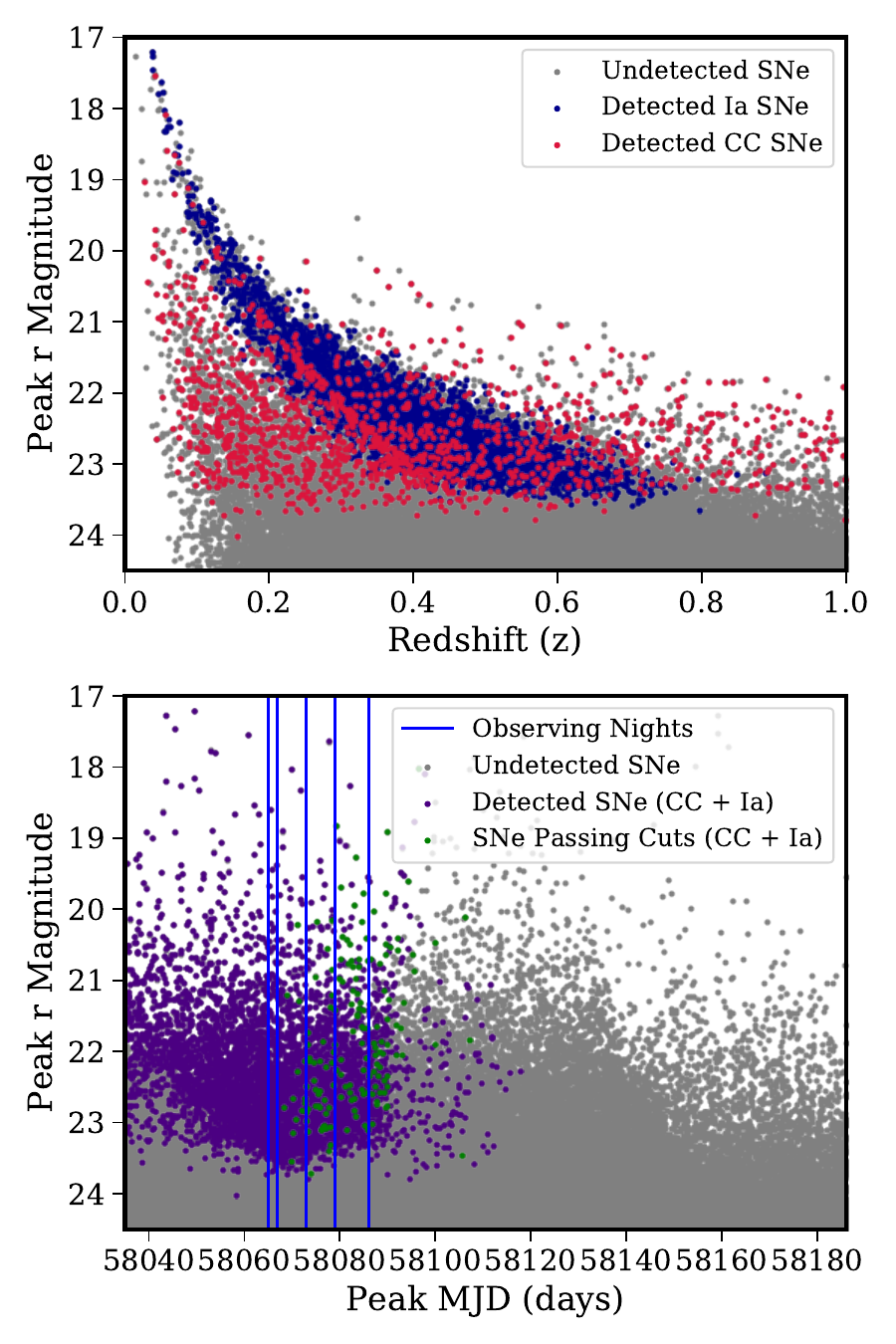}
  \caption{The background SNe distribution for simulations of IC171106A with a limiting magnitude of $23.5 \textrm{ mag}$.}
  \label{fig:ev2_bg_z}
\end{minipage}
\hspace{0.06\textwidth}
\begin{minipage}{.45\textwidth}
  \centering
  \includegraphics[width=\linewidth]{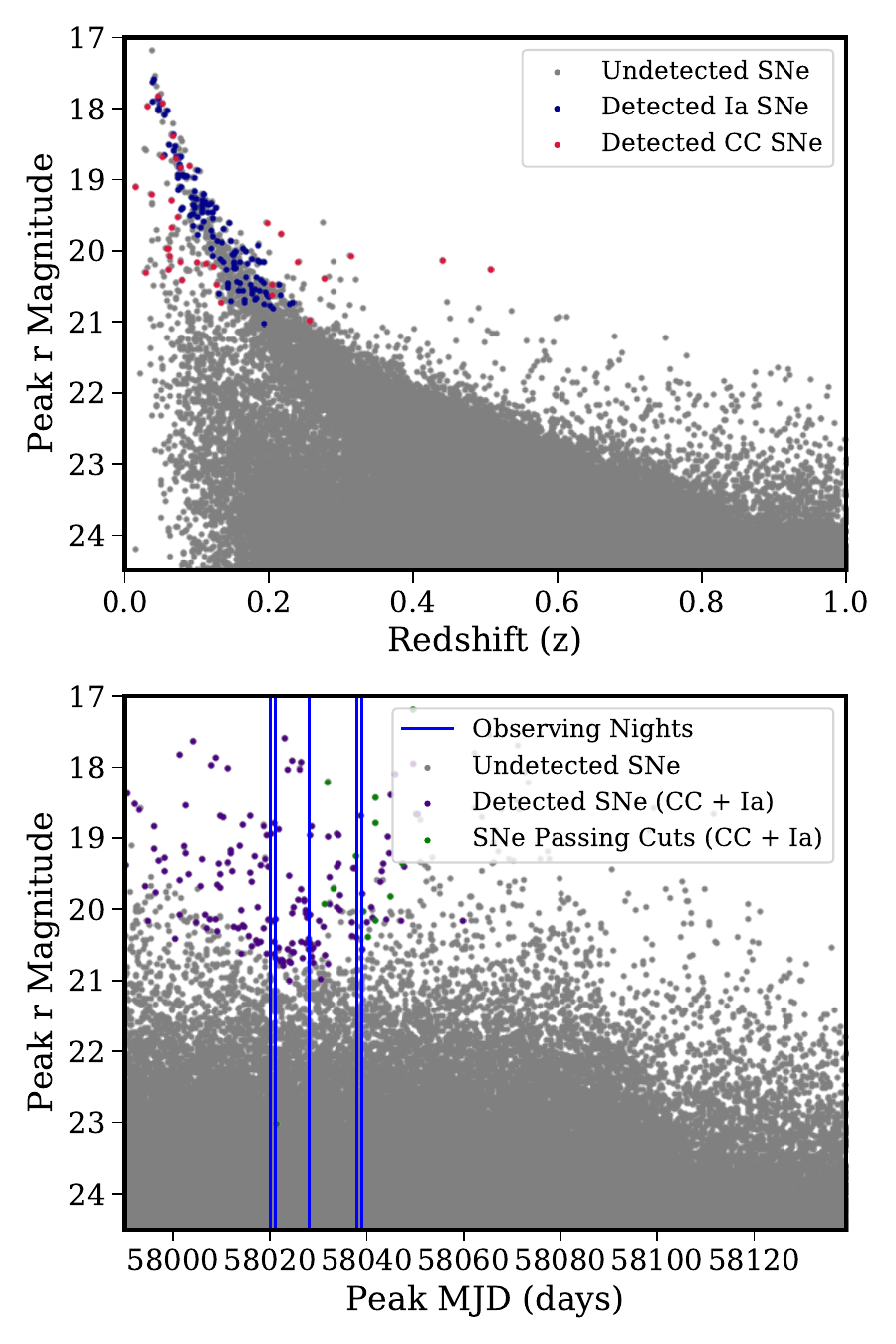}
  \caption{The background SNe distribution for simulations of IC170922A with a limiting magnitude of $21 \textrm{ mag}$.}
  \label{fig:m21_bg}
\end{minipage}%
\end{figure}

\clearpage


\bibliographystyle{aasjournal}
\bibliography{main}

\end{document}